\documentclass[twocolumn,prd,superscriptaddress,nofootinbib,floatfix]{revtex4-1}
\usepackage{graphicx}
\usepackage{natbib}
\usepackage{multirow}
\usepackage[dvipsnames]{xcolor} 
\pdfoutput=1  

\usepackage{graphics,psfrag}
\usepackage{mathrsfs}

\usepackage{algorithm}
\usepackage[noend]{algpseudocode}

\usepackage{amsmath}
\usepackage{amssymb}

\newcounter{magicrownumbers}

\usepackage{soul}

\usepackage[normalem]{ulem} 

\newcommand{\nn}{\nonumber}

\def \be {\begin{equation}}
\def \ee {\end{equation}}
\def \bea {\begin{eqnarray}}
\def \eea {\end{eqnarray}}

\def \ble {\begin{widetext}\begin{equation}}
\def \ele {\end{equation}\end{widetext}}
\def \blea {\begin{widetext}\begin{eqnarray}}
\def \elea {\end{eqnarray}\end{widetext}}

\def \nn {\nonumber}

\newcommand{\eq}[1]{(\ref{#1})}

\def \and {{\textrm{and}}}

\begin{document}
\title{Conditional Noise Deep Learning for Parameter Estimation \\ of Gravitational Wave Events}

\author{Han-Shiang Kuo}
\email{hance30258@gmail.com}
\affiliation{Department of Physics, National Taiwan Normal University, Taipei 11677, Taiwan}

\author{Feng-Li Lin}
\email{fengli.lin@gmail.com, corresponding author.}
\affiliation{Department of Physics, National Taiwan Normal University, Taipei 11677, Taiwan}
\affiliation{Center of Astronomy and Gravitation, National Taiwan Normal University, Taipei 11677, Taiwan}

\date{\today}

\begin{abstract}
We construct a Bayesian inference deep learning machine for parameter estimation of gravitational wave events of binaries of black hole coalescence. The structure of our deep Bayesian machine adopts the conditional variational autoencoder scheme by conditioning on both the gravitational wave strains and the variations of the amplitude spectral density (ASD) of the detector noise. We show that our deep Bayesian machine is capable of yielding posteriors compatible with the ones from the nested sampling method and better than the one without conditioning on the ASD. Our result implies that the process of parameter estimation can be accelerated significantly by deep learning even with large ASD drifting/variation. We also apply our deep Bayesian machine to the LIGO/Virgo O3 events, the result is compatible with the one by the traditional Bayesian inference method for the gravitational wave events with signal-to-noise ratios higher than typical threshold value. We also discuss some possible ways for future improvement.

\end{abstract}
 
\maketitle

\section{Introduction}\label{Introduction}

Detection of gravitational waves (GW) from the distant compact binary coalescence has now become quite common since the first operation runs of LIGO started in 2015 \cite{Abbott:2016blz}, and up to now, about a hundred events have been found \cite{LIGOScientific:2018mvr,Abbott:2020niy}. Due to the extreme weakness of the GW signal, the extraction of the source parameters from a given strain data requires heavy computational cost based on Nested Sampling \cite{skilling2004nested,del2015cpnest,speagle2020dynesty} and  Markov-Chain-Monte-Carlo algorithm \cite{foreman2013emcee,Li:2013lza}, and such task of parameter estimation (PE) is very time-consuming. This will then delay the announcement of the discoveries and the public sharing of the strain data for more general usages and PE results. Once the event rate of detection increases from few events per month to few events per day, this time delay issue of parameter inference will be more severe. Therefore, the acceleration of the PE for GW events is an urgent task in the vision of the improvement of the sensitivity for the new generation of gravitational wave detectors \cite{KAGRA:2020npa}. The main obstacle for accelerating the PE is the time-consuming scan of the likelihood function for obtaining the posteriors in Bayesian inference scheme \cite{PhysRevD.85.122006,Messick_2017,veitch2015parameter,biwer2019pycbc,ashton2019bilby}. One way to bypass this issue is to find a way of performing likelihood-free inference. This is indeed what deep learning can do by training the Bayesian inference machine with lots of mock data so that it can mimic the likelihood without event-by-event scanning. This deep-learning-based machine (or deep machine, for short) can then be implemented to extract the parameters of the GW events in a very efficient way. Some pioneer works in this direction have been done in \cite{gabbard2019bayesian,Green_2020} by adopting the variational autoencoder (VAE) \cite{kingma2019introduction,yu2020tutorial} or the normalizing flow \cite{DBLP:journals/corr/KingmaSW16}, and see \cite{green2021complete,dax2021realtime} for the more recent progress. However, in these works, all training data share the same power spectral density (PSD) (or its square root, the amplitude spectral density (ASD)) of the detector's noise, which may not be realistic since the detector noise will drift in general. This means that the deep machine should be retrained for the events with different ASDs.

In this work, we extend the conditional VAE (CVAE) scheme developed in \cite{gabbard2019bayesian} to also conditioning on the ASD of the detector noise so that the resultant deep machine can deal with the GW events measured at different time intervals, for which the ASD will drift accordingly \footnote{Our machine-learning codes used for this work are implemented based on Tensorflow  \cite{tensorflow2015-whitepaper} and can be found in the open Gitlab forum: https://gitlab.com/hance30258/gwcvae. Besides, we adopt Bilby \cite{ashton2019bilby} to perform conventional Nested Sampling PE dynesty \cite{speagle2020dynesty} to get the results for comparison.}. When finishing this note, we find that a similar consideration is also adopted in recent work \cite{dax2021realtime} in the scheme of normalizing flow. 

The remainder of this paper is organized as follows. In the next section, we will sketch the scheme of CVAE for the inference of the source parameters of the GW events without or with the conditional ASD. In section \ref{section3} we describe how we prepare the training data, especially on how to prepare the variations of ASD and the mock strain data. In section \ref{section4} we describe the detailed structure of our CVAE model, such as the layer structures and the hyperparameters. In section \ref{section5} we first discuss the training procedure, including the way of avoiding KL collapse and the learning rate decay, and carry out the self-check of our Bayesian inference machine. We then show the performance of our machine when applying to the mock data by comparing it to the traditional PE method by their posteriors. We also consider the endurance of our machine to the drift of the ASD when compared to the CVAE model but without conditional ASD. In section \ref{section6}, we apply our Bayesian inference machine to LIGO/Virgo O3 events \cite{vallisneri2015ligo,gracedbO3}, and show their performance. Finally, we conclude our paper in section \ref{conclusion}.

\section{CVAE for Bayesian Inference of GW Events}\label{section:cVAE}

\begin{figure}[htp]
	\centering\resizebox{8cm}{!}{\includegraphics{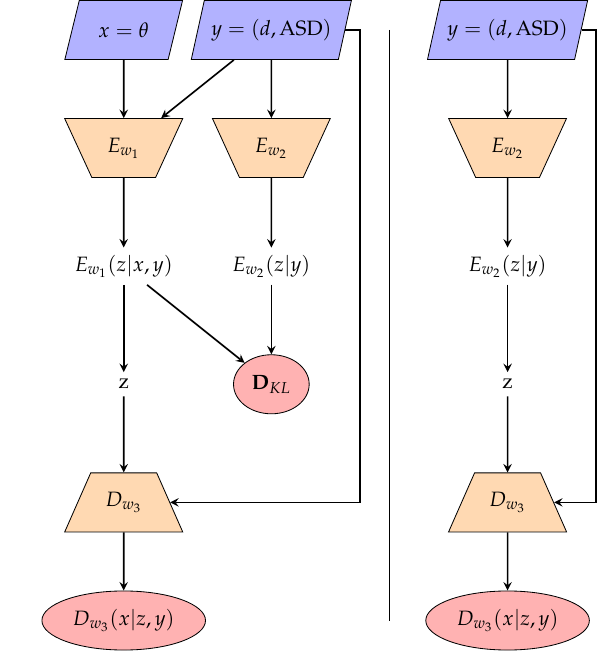}}
\caption{The schematic structure of CVAE for the inference of source parameters of GW events. The goal is to generate the posteriors $p(x|y)$ of source parameters efficiently for a given strain data without knowing the likelihood $p(y|x)$. {\bf (Left)} the CVAE machine with two encoders $E_{w_1}$, $E_{w_2}$ and one decoder $D_{w_3}$. {\bf (Right)} the Bayesian inference machine, which is obtained by removing the $E_{w_1}$ part of CVAE after the CVAE is well-trained, so that we expect $p(x|y) \approx {\bf E}_{z\sim E_{w_2}(z|y)}[D_{w_3}(x|z,y)]$. Therefore, its outputs are the posteriors of the source parameters. Our scheme shown here is a generalization of \cite{gabbard2019bayesian} by adding the ASD of the detector noise as the conditional inputs besides the associated strain data.} \label{CVAE-PE}
\end{figure}

The variational autoencoder (VAE) is a unsupervised machine learning scheme, which can be used to reveal the distribution functions of the input data. It first compresses the input data into the hidden layer by its encoder part, and then decompresses the hidden layer into the output by its decoder part. For example, if we prepare many mock strains as the training data, then the resultant well-trained machine can learn the distribution of the strains, and the hidden layers will encode the information about the distribution of the source parameters. However, to make the VAE be useful for the inference of the source parameters, we need to train the machine by simultaneously providing the strains $\{ y \}$ and the associated source parameters $\{ x\}$ as the input data but feeding to different encoders. The schematic structure of CVAE is similar to what is shown in Fig. \ref{CVAE-PE}. The loss function of this machine can be thought of as the upper bound on the negative of the posterior distribution $p(x|y)$, i.e., the so-called evidence lower bound (ELBO) and denoted by ${\cal L}_{\textrm{ELBO}}$, 
\bea
 -\log p(x|y) &\le & {\bf E}_{z\sim E_{w_1}(z|x,y)} [-\log D_{w_3}(x|y,z)] \qquad
 \nn\\ \label{CVAE-loss}
 && + {\bf D}_{KL}[E_{w_1}(z|x,y)||E_{w_2}(z|y)] 
\eea
where $E_{w_i}$ for $i=1,2$ denote the distributions of the encoders with the associated weights and biases denoted by $w_i$, and $D_{w_3}$ the one of the decoder with $w_3$ the associated weights and biases. Moreover, the arguments and the conditional arguments of the encoders and decoder denote their outputs and inputs, respectively.  The right-handed side of the first line of \eq{CVAE-loss} is the so-called reconstruction loss measuring the difference between input and output, and the second line is the Kullback–Leibler (KL) loss measuring the difference between the hidden layer distributions of the two encoders.

After the training, we can remove the part associated with the source parameters but keep only the one associated with the strains, so that the remaining part (as shown on the right part of Fig. \ref{CVAE-PE}) can be treated as the Bayesian inference machine to output the posteriors of the source parameters for a given input strain. Namely, we expect 
\be\label{posterior}
p(x|y) \approx {\bf E}_{z\sim E_{w_2}(z|y)}[D_{w_3}(x|z,y)]\;.
\ee
Even though $D_{w_3}(x|z,y)$ is a Gaussian distribution, the average over $z\sim E_{w_2}$ will lead to non-Gaussian posterior approximation, as generally expected. 

The above scheme was first proposed and implemented in \cite{gabbard2019bayesian}, and can be shown to produce compatible posteriors in comparison to the conventional PE. However, in reality, the ASD of the detector's noise can drift so that ASD varies event by event. This drifting effect has not been taken into account in \cite{gabbard2019bayesian}. In this work, we extend the CVAE scheme of \cite{gabbard2019bayesian} to also include the variations of ASD as the conditional input data. The new scheme is shown in Fig. \ref{CVAE-PE}.  This is the same as the one implemented in \cite{gabbard2019bayesian} except that an ensemble of ASD is also conditioned when training, and an ASD should be provided as the input along with the corresponding strain data when generating the posteriors of a GW event by the resultant Bayesian inference machine, i.e., the right part of the Fig. \ref{CVAE-PE}.

\section{Preparation of training data}\label{section3}
As discussed, the training data include both the strain data and the ASD of the detector noise.  As a proof-of-concept study, we only consider the binaries of black holes (BBH) without spin, which are labeled by two intrinsic parameters, i.e., the component masses $m_1$, $m_2$. Besides, we also have the extrinsic parameters describing the locations of the binaries. For simplicity, we fix all the extrinsic parameters except the luminosity distance $d_L$, which dictates the signal-to-noise ratio (SNR). Moreover, in the usual conventional PE, we need to optimize the matched filtering overlap by adjusting the time of coalescence $t_c$ and phase at coalescence $\phi_0$. Thus, we also include  $t_c$ and $\phi_0$ as the parameters for inference. In total, we have five parameters for inference, and their ranges for flat priors and the fixed values of other parameters are given in Table \ref{table:prior}. 

\begin{table}[htb]
\caption{Ranges of the priors for the BBH GW events adopted for the training data of the CVAE models used in this paper.}\label{table:prior}
\vspace*{3mm}
\centering
\resizebox{\columnwidth}{!}{%
\begin{tabular}{ccccc}
\hline
\hline
parameters& symbol& prior&  $\operatorname{range}^a$& units\\
\hline
mass 1& $m_1$&  Uniform&    [20, 65]&    solar masses\\
mass 2& $m_2$&  Uniform&    [20, 65]&    solar masses\\
luminosity distance&    $d_L$&  Uniform Volume&  [1200, 2200]&     Mpc\\
time of coalescence&    $t_c$ &  Uniform&        [0.65, 0.85]&    seconds\\ 
phase at coalescence&   $\phi_0$&   Uniform&    [0, 2$\pi$]& radians\\
\hline
right ascension&    $\alpha$&   $\cdot$&    1.84&       radians\\
declination&        $\delta$&   $\cdot$&    -0.62&      radians\\
inclination&        $\eta$&     $\cdot$&    0&  radians\\
polarization&       $\phi$&     $\cdot$&    0&  radians\\
epoch&              $\cdot$&    $\cdot$&    1242459857& GPS time\\
detector&           $\cdot$&    $\cdot$&    Livingston& $\cdot$\\
\hline                              					\hline
\end{tabular}%
}
\vspace*{3mm}
\par
\footnotesize{\raggedright 
$^a$ The prior ranges chosen here are aiming at performing the PE of the real LIGO/Virgo's O3a events with some adjustments due to the limitation of our computing resources. 
\par}
\end{table}

Unlike in \cite{gabbard2019bayesian}, we adopt the frequency-domain templates to generate the strain data of one-second duration, instead of the time-domain ones as in \cite{gabbard2019bayesian}. We also change the sampling rate from $256$Hz to $1024$Hz to yield waveforms of $512$Hz bandwidth after fast Fourier transform and cover the high-frequency part of typical real-data waveforms.  With the setup of priors given in Table \ref{table:prior}, we sample $2\times 10^6$ sets of parameters to produce the theoretical waveforms by the IMRPhnomPv2 waveform model \cite{khan2019phenomenological}, which later will be used to superpose with the sample detector noise to produce the mock strain data. 

Now, we turn to the preparation of the set of ASD templates used for training our CVAE. To take care of the ASD variations in LIGO/Virgo detectors, we would like to simulate an ASD model by fitting to a backbone set of 2000 sample ASDs, each of which is obtained from a 4096-second segment of the O3a strain data. These 2000 segments released by LIGO Scientific Collaboration cover the whole strain data of LIGO/Virgo's O3a run-time. Each 4096-second segment is divided into 4096 one-second pieces. Each piece is sampled with 1024Hz to get a component ASD. The mean of the 4096 component ASDs yields a sample ASD in the backbone set. A 4096-second segment seems too long for a typical one-second BBH event due to significant noise-drifting. However, by taking the mean to yield an ASD, the glitches could be averaged out. Since our CVAE model is trained to combat ASD variations, thus we choose such a long segment to eliminate the glitches in advance.

To construct a stochastic ASD model to simulate ASD variations from the above backbone set of ASDs, we first construct a minimal ASD profile $A_m[f]$ obtained by collecting the minimum of each frequency bin from the above backbone set of ASDs. Based on $A_m[f]$ as shown in Fig. \ref{fig:asd_comparison}, we assume that the variations of the logarithm of the ASD $A[f]$ mainly obey the chi-square distribution $\alpha \equiv Z_1^2+Z_2^2 $ of two degrees of freedom with $Z_{1,2}=\mathcal{N}(0,1)$,  but with a small but uniform residual variation $\beta \equiv \mathcal{N}(0,1/8)$ for all frequency bins. Noting that $\mathcal{N}(\mu,\sigma)$ denotes a Gaussian distribution with mean and variance $(\mu,\sigma^2)$. Thus, our ansatz stochastic ASD model is given by 
\be\label{ASD_o3a}
\delta \log{A[f]}\equiv \log{A[f]} -\log{A_m[f]} =\alpha \log{r[f]} + \beta\;.
\ee 
We then use the aforementioned backbone set of ASDs to fit the overall variation factor $r[f]$ of 513 frequency bins, see more discussion of the fitting procedure in Appendix \ref{fitting}. In Fig. \ref{fig:chi2}, we show the chi-square distribution $\alpha$ and the corresponding confidence level,  and in Fig. \ref{fig:drifting_factor} we show the overall variation factor $r[f]$ which turns out to be almost a continuous function, and the histograms of the overall drifting factor $\Big(r[f]\Big)^\alpha$. We see that the overall ASD variation factor can be larger than $15$ for some frequency ranges at the $90\%$ confidence level of the ASD variation.

Since our ASD model \eq{ASD_o3a} is obtained by fitting the sample ASDs of the whole LIGO/Virgo's O3a run-time, thus the ASDs generated from it can catch the generic features of ASD variation/drifting during the O3a. In Fig. \ref{fig:asd_comparison}, we show the range of the ASD variations generated by the stochastic ASD model \eq{ASD_o3a}.
However, we should also caution the readers that the task of constructing a perfect ASD model is almost impossible since not all origins of the variations are well understood, e.g., the unexpected frequent glitches. Our ASD model \eq{ASD_o3a} is simple and may not catch the full features of the ASD variations.

\begin{figure}[ht]
	\centering\resizebox{8cm}{!}{\includegraphics{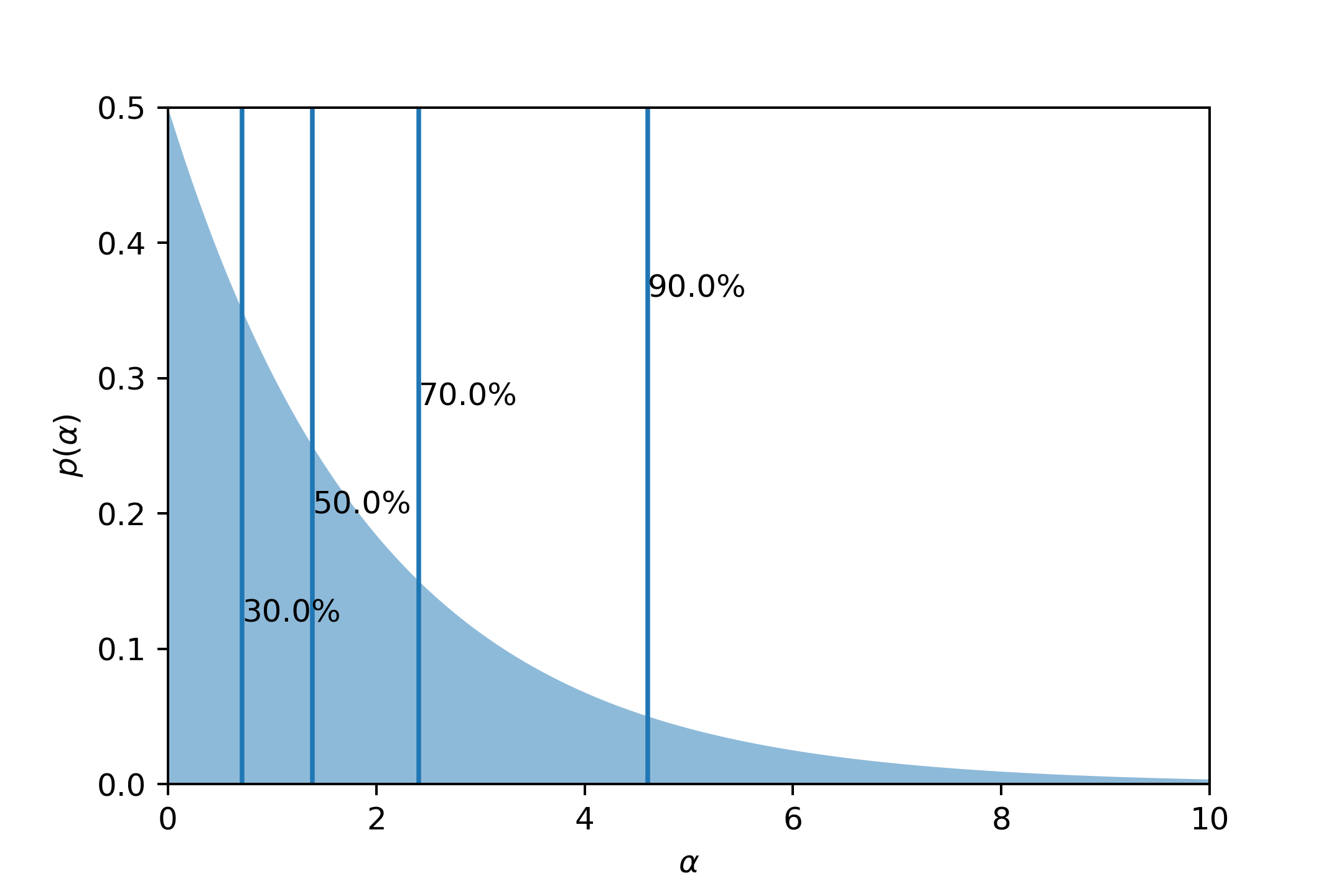}}
	\caption{Chi-square distribution for $\alpha$ used in \eq{ASD_o3a}, and the vertical lines indicate the corresponding confidence levels.} \label{fig:chi2}
\end{figure}
\begin{figure}[ht]
	\centering\resizebox{8cm}{!}{\includegraphics{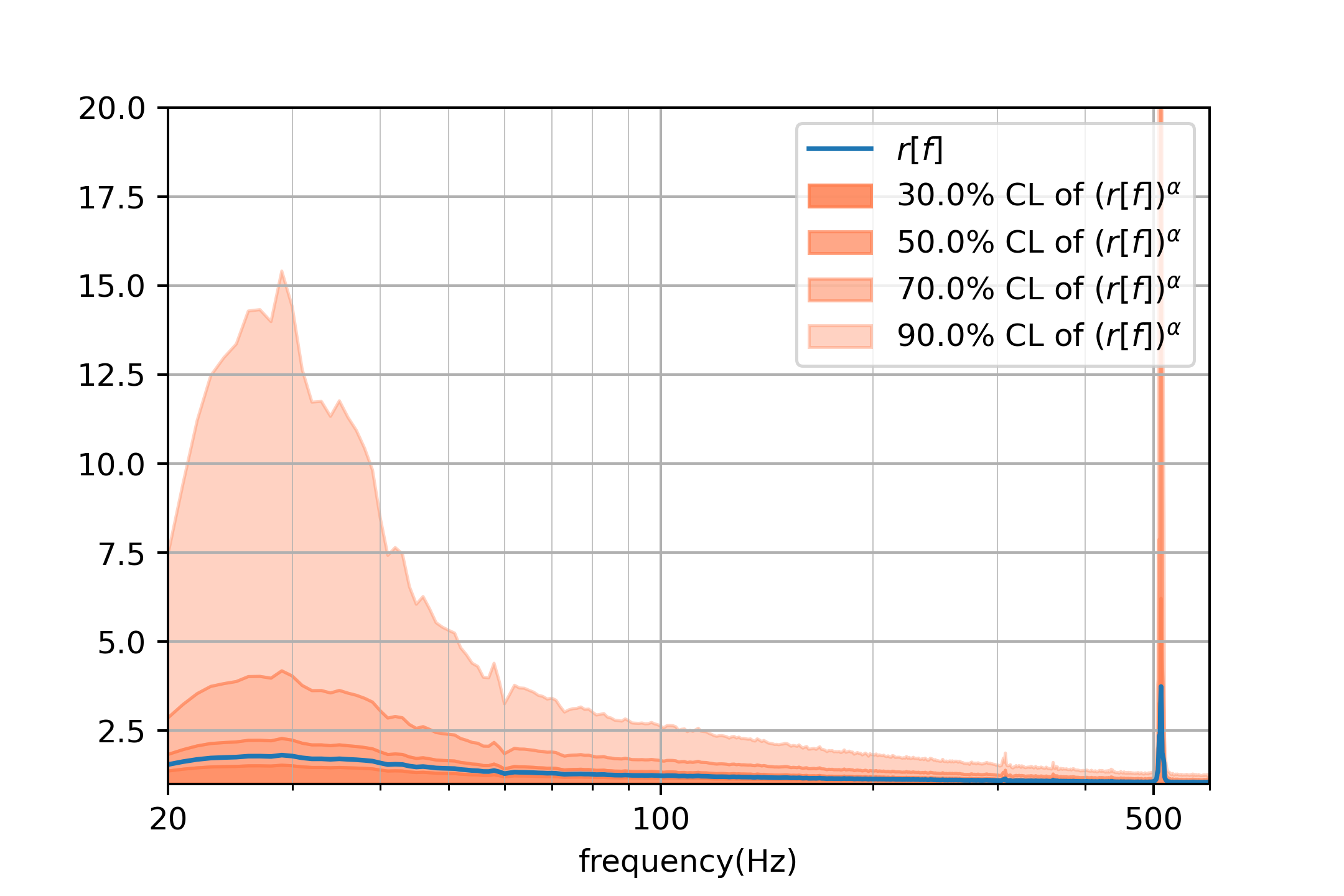}}
	\caption{Profiles of ASD variations from the stochastic ASD model \eq{ASD_o3a}. The blue curve is the fitted $r[f]$. The shaded pink regions indicate the various profiles of the overall variation factor $(r[f])^\alpha$ at $30\%$, $50\%$, $70\%$ and $90\%$ confidence levels, respectively. We see that the overall ASD variation factor can be larger than 15 for some frequency range.} \label{fig:drifting_factor}
\end{figure}

We will use the ASD model \eq{ASD_o3a} to generate about two million ASDs as the training set for our CVAE model. Therefore, the resultant PE model will learn the generic feature of ASDs and can endure the drifting of ASD.  This contrasts with the PE model considered in \cite{gabbard2019bayesian,Green_2020}, in which the ASD used to whiten the strain data is fixed so that the model should be retrained if the drifting of the ASD is significant.  

\begin{figure}[ht]
	\centering\resizebox{8cm}{!}{\includegraphics{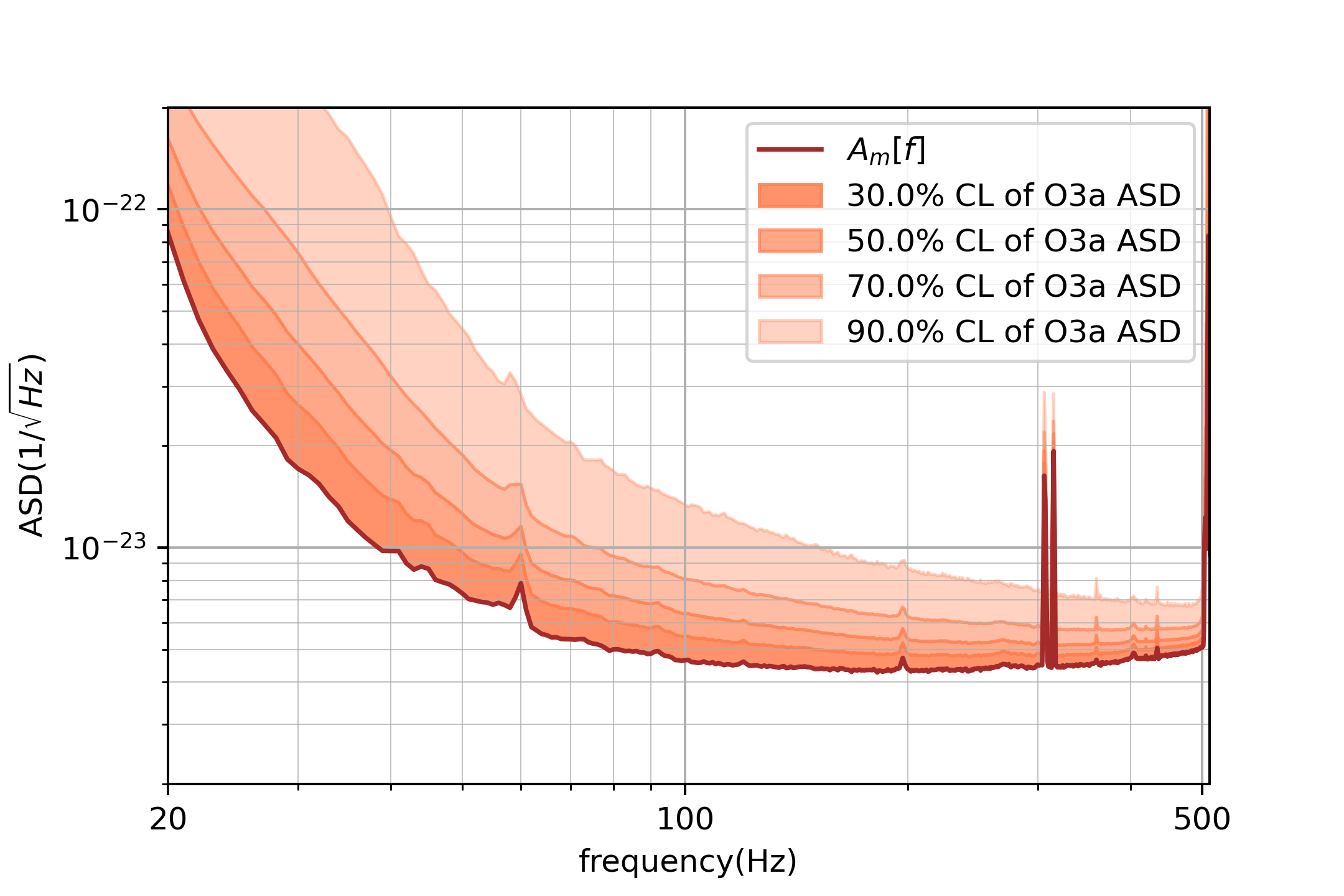}}
	\caption{Ranges of ASDs generated from our stochastic ASD model \eq{ASD_o3a} with the pink shaded regions indicating the profile ranges at $30\%$, $50\%$, $70\%$ and $90\%$ confidence levels, respectively. The brown curve is the minimal ASD profile $A_m[f]$. } \label{fig:asd_comparison}
\end{figure}

\begin{figure}[ht]
	\centering\resizebox{8cm}{!}{\includegraphics{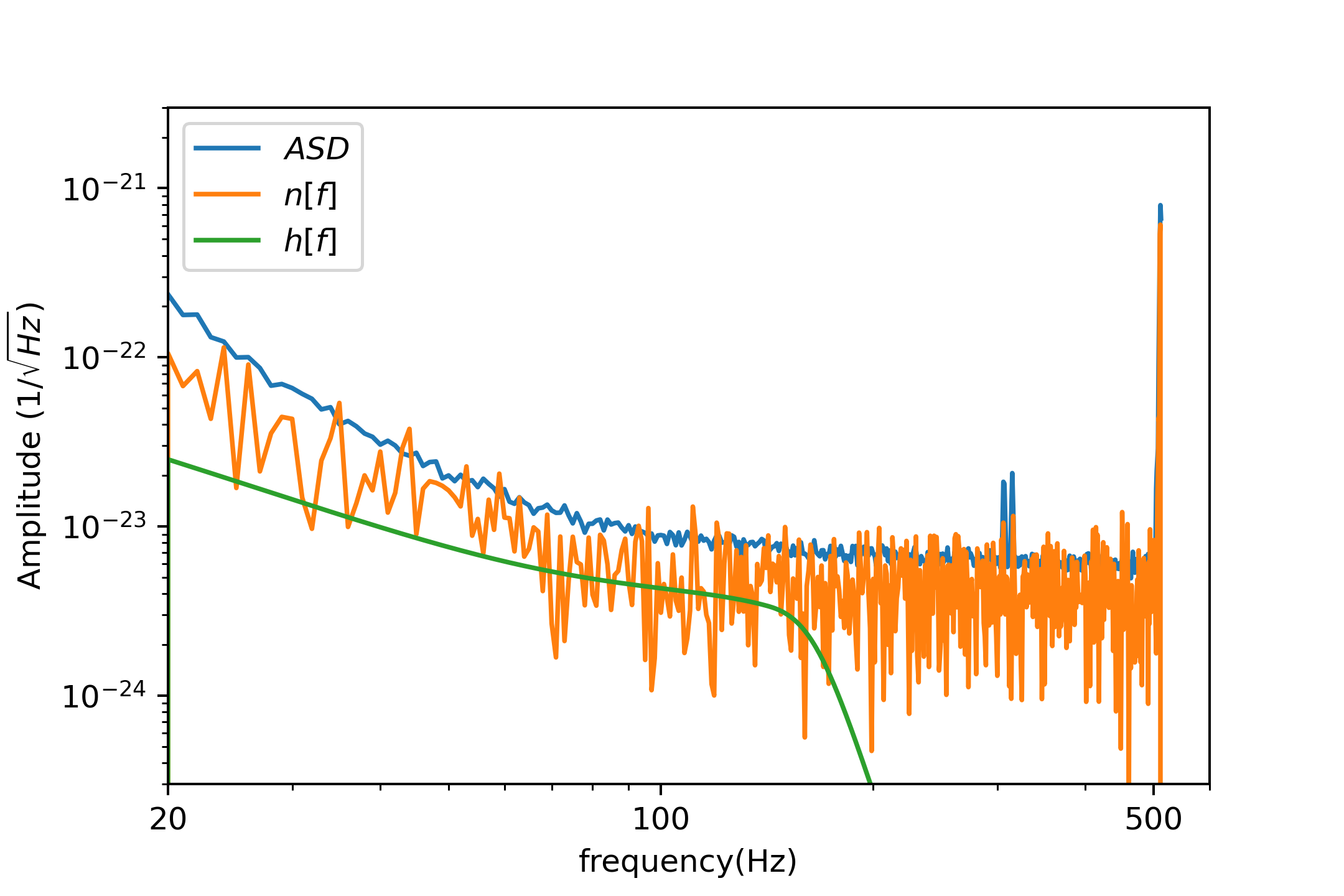}}
	\caption{A typical mock waveform $h[f]$ (green) and noise $n[f]$ (orange). The latter is generated from a given ASD (blue) obtained by \eq{ASD_o3a}. Here we only plot the amplitude part. }\label{fig:strain}
\end{figure}

Based on the above discussions, we can generate a mock strain as follows. We randomly pick up a theoretical waveform $h[f]$ and ASD $A[f]$ from the above prepared sets, then we can form a noise $n[f]$ and a strain $d[f]$ in the frequency domain as follows
\bea
    d[f] &=& h[f] + n[f],\\
    n[f] &=& \frac{1}{\Delta f} W[f] \odot A[f]
\eea
where $\Delta f$ is the frequency bin size which we set to $1$Hz in this work, and $W[f]$ is the white noise in the frequency domain, which is responsible for the unit Gaussian noise. In Fig. \ref{fig:strain}, we show a typical example.

With the above procedure, we generate about $2\times10^6$ mock strains, of which $80\%$ will be used as the training data set for CVAE, and $20\%$ as validation data set for the resultant Bayesian inference machine. This amount of the training data set is huge enough to exhaust almost all possible strain data realizations. 

\begin{figure}[ht]
	\centering\resizebox{7cm}{!}{\includegraphics{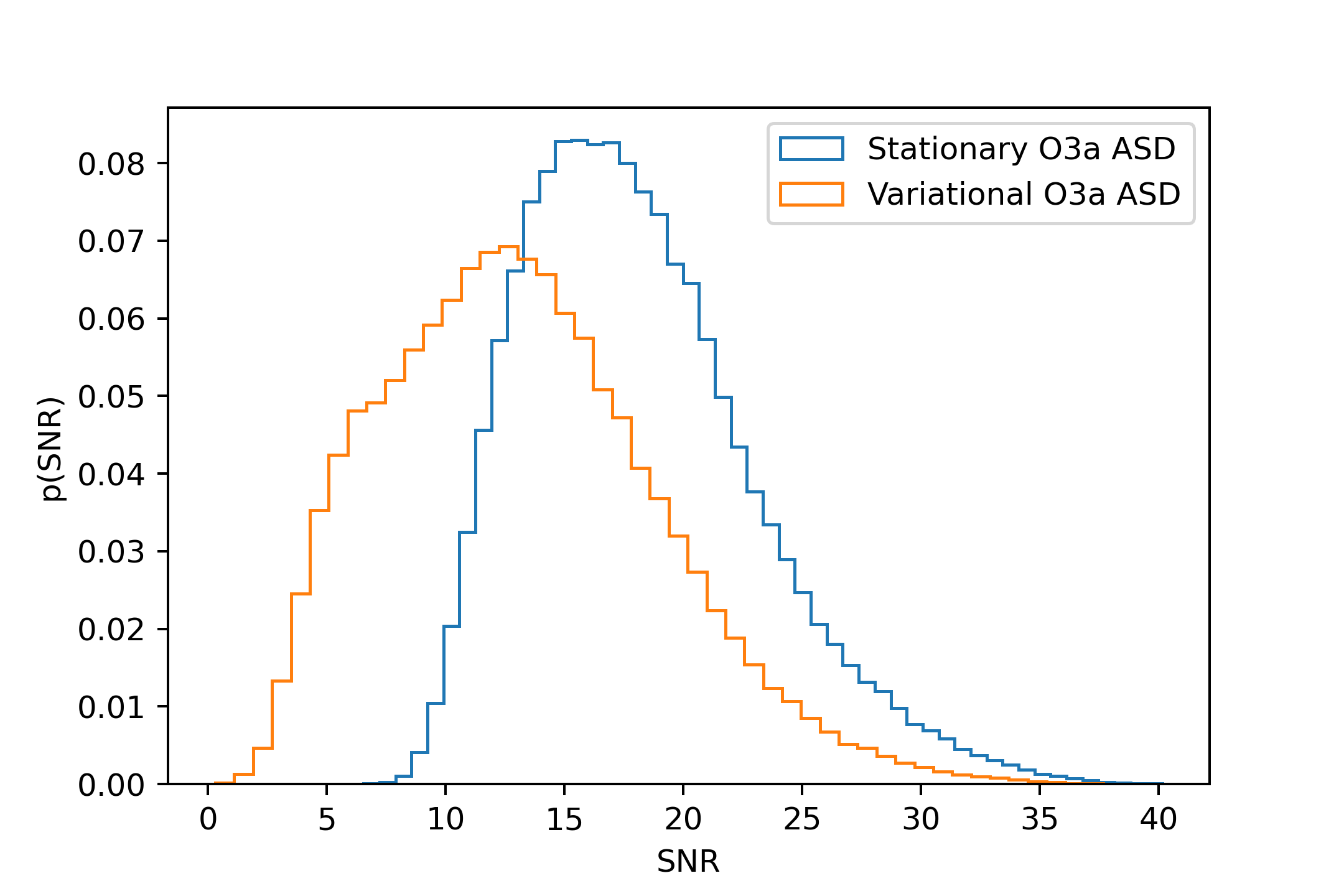}}
	\caption{Histograms of SNRs for all the training strain data generated by using the priors in Table \ref{table:prior} and the ASDs generated by $A_m[f]$ (blue) and $A[f]$ (orange) of \eq{ASD_o3a}.
	It shows a significant down-shift of the SNR distribution due to the noise-drifting.}\label{fig:snr}
\end{figure}

Moreover, to quantify how the ASD variations affect the quality of the strain data, we compare the histograms of the SNR obtained from the $2\times 10^6$ mock strain data with and without ASD variation. The result is shown in Fig. \ref{fig:snr}, from which we can see that the SNR distribution is downshifted by quite an amount. This is expected, as the drifting factor at some particular frequency range can be as large as $15$. It implies that a well-trained CVAE model augmented by our ASD model \eq{ASD_o3a} should be able to combat the noise drifting and yield reliable PE results.

\section{The detailed structure of CVAE model}\label{section4}
\begin{figure}[!htb]
	\centering\resizebox{6cm}{!}{\includegraphics{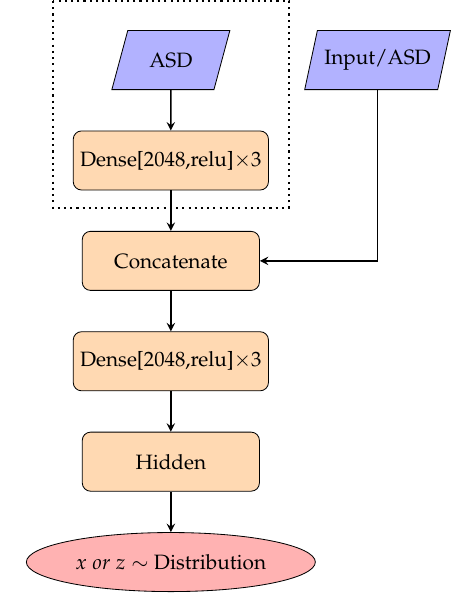}}
	\caption{Structure of neural network used in CVAE model of Fig. \ref{CVAE-PE}. Note that we adopt this same NN for all three NNs, i.e., $E_{w_1}$, $E_{w_2}$ and $D_{w_3}$ of Fig. \ref{CVAE-PE}. The ASD is the common input, and there is an additional input denoted by $\textrm{Input/ASD}$. The $\textrm{Input/ASD}$ and the dimensions of the hidden layer vary for different NNs, which we summarize in Table \ref{table:cvaep_nn}. The output is a random latent vector, $z  \mbox{ or } x \sim \textrm{Distribution}$ which is also specified in Table \ref{table:cvaep_nn}.  The dash-lined box contains the part associated with the conditional ASD, which is absent in the CVAE model of \cite{gabbard2019bayesian}.
	}\label{fig:common_nn}
\end{figure}

The schematic structure of our CVAE model and the resultant Bayesian inference machine has been shown in Fig. \ref{CVAE-PE}. Now we would like to expose its detailed structure. For simplicity, our CVAE model is composed of only dense layers but not other types of layers. However, it works. We simply stack the dense layers to construct three neural networks (NNs), i.e., two encoders and a decoder. Moreover, we adopt almost the same layer structure for all three NNs, see Fig. \ref{fig:common_nn} for the details. The only differences among them are the input data and the dimensions and the realizations of the hidden layers. Specifically, we use 8- and 5-dimensional multivariate Gaussian distributions for the hidden layers of $E_{w_1}$ and $D_{w_3}$, and adopt a more powerful mixture Gaussian distribution layer for $E_{w_2}$, which has eight dimensions and each dimension is composed of eight components of Gaussian normal distributions. From our experience, adopting the mixture Gaussian for the hidden layer enhances a lot the performance of the model. All hidden layers with Gaussian distributions are realized by the standard reparameterization trick used for variational autoencoder \cite{kingma2019introduction}. In Table \ref{table:cvaep_nn}, we summarize these differences.

Note that the latent vectors for the encoders $E_{w_1}$ and $E_{w_2}$ are denoted by $z$, which will then be input to the decoder. However, the output of the decoder is again a random vector, whose components are identified as the source parameters, i.e., $x=\theta$. The distribution of $x$ gives the approximate posterior of the source parameter $\theta$ through the averaging procedure given in \eq{posterior}.

\begin{table}[!htb]
	\caption{Input and Hidden Layers of the CVAE model} \label{table:cvaep_nn}
	\vspace*{3mm}
	\setcounter{magicrownumbers}{0}
	\centering
	\resizebox{\columnwidth}{!}{%
	\begin{tabular}{cccc}
		\hline
		\hline
		 & $E_{w_1}(z|x,y)^a$ & $E_{w_2}(z|y)$ & $D_{w_3}(x|z,y)$\\
		\hline
		$\textrm{Input/ASD}^b$ & $[\theta,d]$ & $d$ &	$[z,d]$\\
		$\textrm{Hidden}^c$ & $\operatorname{Dense}[16,\mathsf{linear}]$ & $\operatorname{Dense}[24,\mathsf{linear}]$ &	$\operatorname{Dense}[10,\mathsf{linear}]$ \\
		$\textrm{Distribution}^d$ &	${\it Gaussian}(8)$ & ${\it MixtureNormal}(8,8)^e $ & ${\it Gaussian}(5)$ \\
		
		\hline
		\hline
	\end{tabular}%
	}
	\vspace*{3mm}
	\par
	\footnotesize{\raggedright
	$^a$ Here $x=\theta$ denoting the source parameters, $y=(\textrm{ASD},d)$ with $d$ the strain, and $z$ the random latent vector. 
	
	$^b$ This means the additional input other than ASD. 
	
	$^c$ This is the hidden layer whose outputs are means and variances. 
	
    $^d$ This is the distribution used to generate the random latent vector $z$, whose means and variances are given by the outputs of the hidden layers.
    
	$^e$ This is the linear combination of 8 Gaussian distributions.
	\par}
\end{table}

The hyperparameters specified in Fig.\ref{fig:common_nn} and Table \ref{table:cvaep_nn} achieve well-training of our CVAE model, despite that they can be varied. However, we find that it is sufficient for well-training if the dimension of the hidden layer is greater than the dimension of the target variable. Using more dimensions may need a longer time to train but improve the performance just slightly.

\section{Training the CVAE Model and the Performance}\label{section5}
With the above structure of the CVAE model, we train the model by the aforementioned training data set of batch size 2048. We then calculate the loss function, i.e., ${\cal L}_{\textrm{ELBO}}$ and update the model by Adam optimizer \cite{kingma2014adam} with learning rate $10^{-4}$. The reconstruction loss is evaluated by replacing $x$ in ${\bf E}_{z\sim E_{w_1}(z|x,y)} [-\log D_{w_3}(x|y,z)]$ by the input source parameter $\theta$. The averaging procedure over $z$ in the above and in evaluating the $KL$ loss is done by the  Monte-Carlo method.
There are two effective ways to achieve well-training. The first way is addressed to the so-called KL collapse \cite{lucas2019understanding}, which states that KL loss may happen to be extremely small so that the variational nature of CVAE is lost. To avoid the KL collapse, we can adopt the annealing procedure by introducing an annealing factor $b \in [0,1]$ so that the ELBO is changed to 
\be 
\mathcal{L}^{(b)}_{\textrm{ELBO}} = {\bf E}_{z\sim E_{w_1}}[-\log D_{w_3}] + b \; {\bf D}_{\textrm{KL}}[E_{w_1}||E_{w_2}] 
\ee
In the early training phase, we slowly tune up the annealing factor to avoid the KL collapse. When $b$ is far smaller than one, we are mainly training the VAE, i.e., ignoring the $E_{w_2}$ which will be optimized again when $b$ is close to one. Specifically, we proceed the KL annealing for the first 5 epochs with the following annealing behavior
\be
b(t) = b_0\; \sin(\frac{\pi}{2}t/c),
\ee
where $t$ denotes the number of generations (each generation means finishing a batch training) and $c \approx 10^3$ is the number of generations within an epoch, and the values of $b_0$ for these 5 epochs are set to $[ 10^{-2}, \frac{1}{4}, \frac{1}{2}, 1, 1]$. Note that the annealing rate gradually approaches zero at the end of each epoch. After these 5 epochs, $b$ will be set to one for the remaining training period, which is about $10^3$ epochs.

\begin{figure}[!htb]
 \centering\resizebox{8cm}{!}{\includegraphics{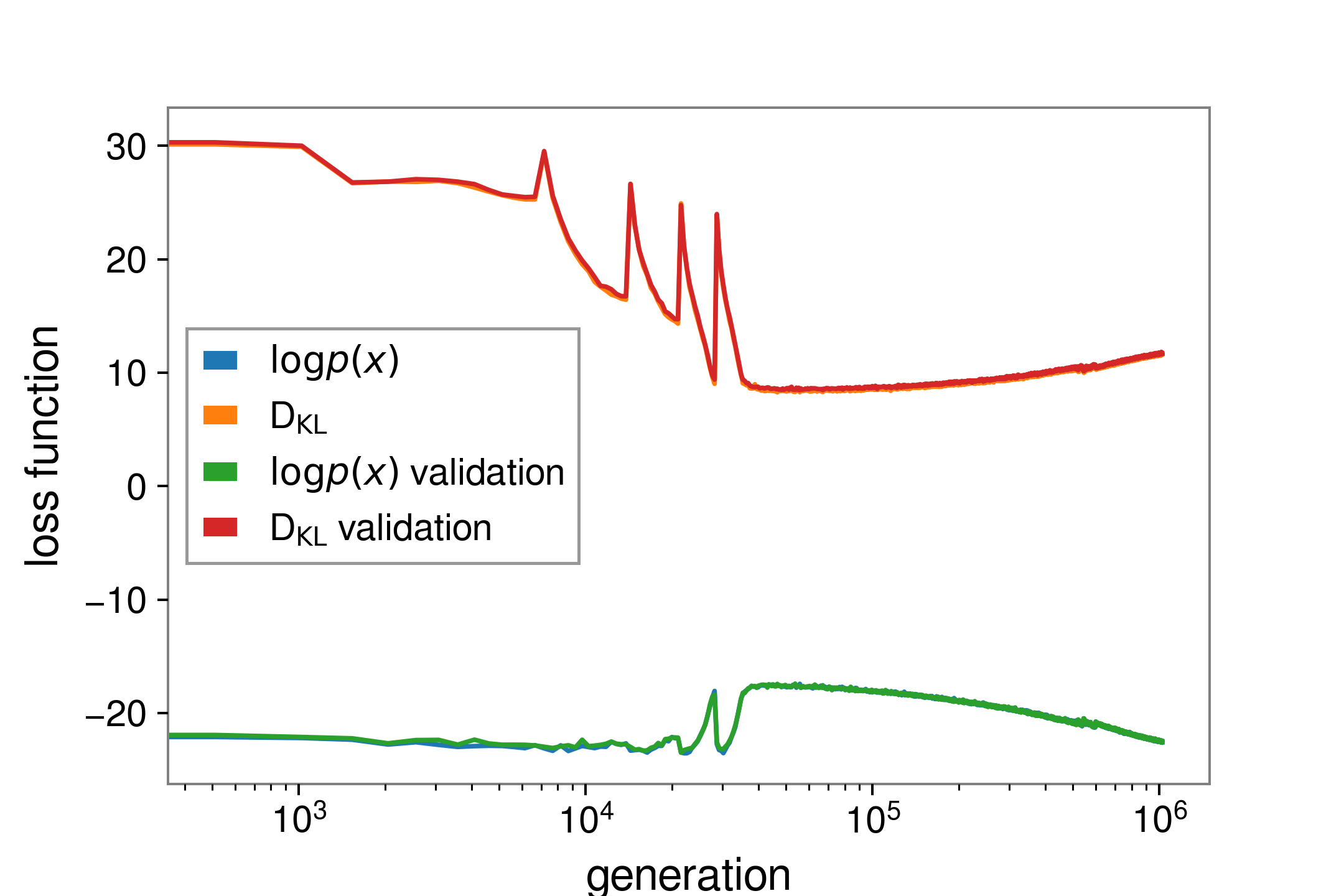}}
  \caption{Training and validation loss for each generation. The variation at early stage are caused by cyclic KL annealing \cite{fu2019cyclical}. The perfect overlap between training and validation loss indicates there is no overfitting.} \label{fig:loss}
\end{figure}

The second effective way to achieve well-training more efficiently is to reduce the learning rate gradually. We reduce the learning rate $\textrm{lr}$ at every generation at such a rate $\textrm{lr}(t)=2^{-{t\over 2\times 10^5}} \; \textrm{lr}_0$ in the total training period of $10^6$ generation. With the implementation of the above two effective ways, we can achieve well-training of our CVAE model. A typical example for the evolution of the reconstruction and KL losses at the training and validation periods is shown in Fig. \ref{fig:loss}, which indicates the KL annealing at the early training phase. Moreover, the perfect overlap between training and validation losses indicates there is no overfitting.

In the following, we will compare our conditional-ASD CVAE model, which we denote as $\overline{\textrm{CVAE}}_{\textrm{ASD}}$, and the one used in \cite{gabbard2019bayesian} but with KL annealing and learning rate decay incorporated, which we denote as $\overline{\textrm{CVAE}}_{\textrm{nc-ASD}}$ with ``no-conditioning" short-handed by $\textrm{nc}$. That is, $\overline{\textrm{CVAE}}_{\textrm{nc-ASD}}$ has the same layer structure as shown in Fig. \ref{fig:common_nn} except for the part inside the dash-lined box, which is used for conditioning on ASD \footnote{This is the layer structure used in the version 1 and 2 of \cite{gabbard2019bayesian}. In the latest version (version 3) of \cite{gabbard2019bayesian}, a more complicated structure with convolutional neural networks is adopted. However, the performance of P-P plot \cite{ghasemi2012normality} and KL divergence \cite{perez2008kullback} of $\overline{\textrm{CVAE}}_{\textrm{nc-ASD}}$ shown below is still better than the latest ones in \cite{gabbard2019bayesian}.}.
Note that we implement  $\overline{\textrm{CVAE}}_{\textrm{nc-ASD}}$ by our own code and then train it with strain data whitened by stationary ASD, i.e., the $A_m[f]$ given in \eq{ASD_o3a}. Also, the mixture Gaussian distribution is used for $E_{w_2}$ in $\overline{\textrm{CVAE}}_{\textrm{nc-ASD}}$ rather than the simple diagonal Gaussian distribution used in \cite{gabbard2019bayesian}. Here, the over-line is to remind that the KL annealing and learning rate decay are implemented in the training procedure. This is in contrast to the CVAE model used in \cite{gabbard2019bayesian}, which we denote as $\textrm{CVAE}_{\textrm{nc-ASD}}$. It turns out that the implementation of KL annealing and learning rate decay in the training procedure is important in achieving better accuracy of final posteriors, as shown below in comparing the P-P plots and histograms of KL divergences.   

Our models are all trained by one GPU device, NVIDIA RTX3090, where the training time is 12 hours for $\overline{\textrm{CVAE}}_{\textrm{ASD}}$ and 6 hours for $\overline{\textrm{CVAE}}_{\textrm{nc-ASD}}$. However, the computational times for evaluating $10^5$ distribution samples for these two models are all below 1 second.

\begin{figure}[!htb]
	\centering\resizebox{8cm}{!}{\includegraphics{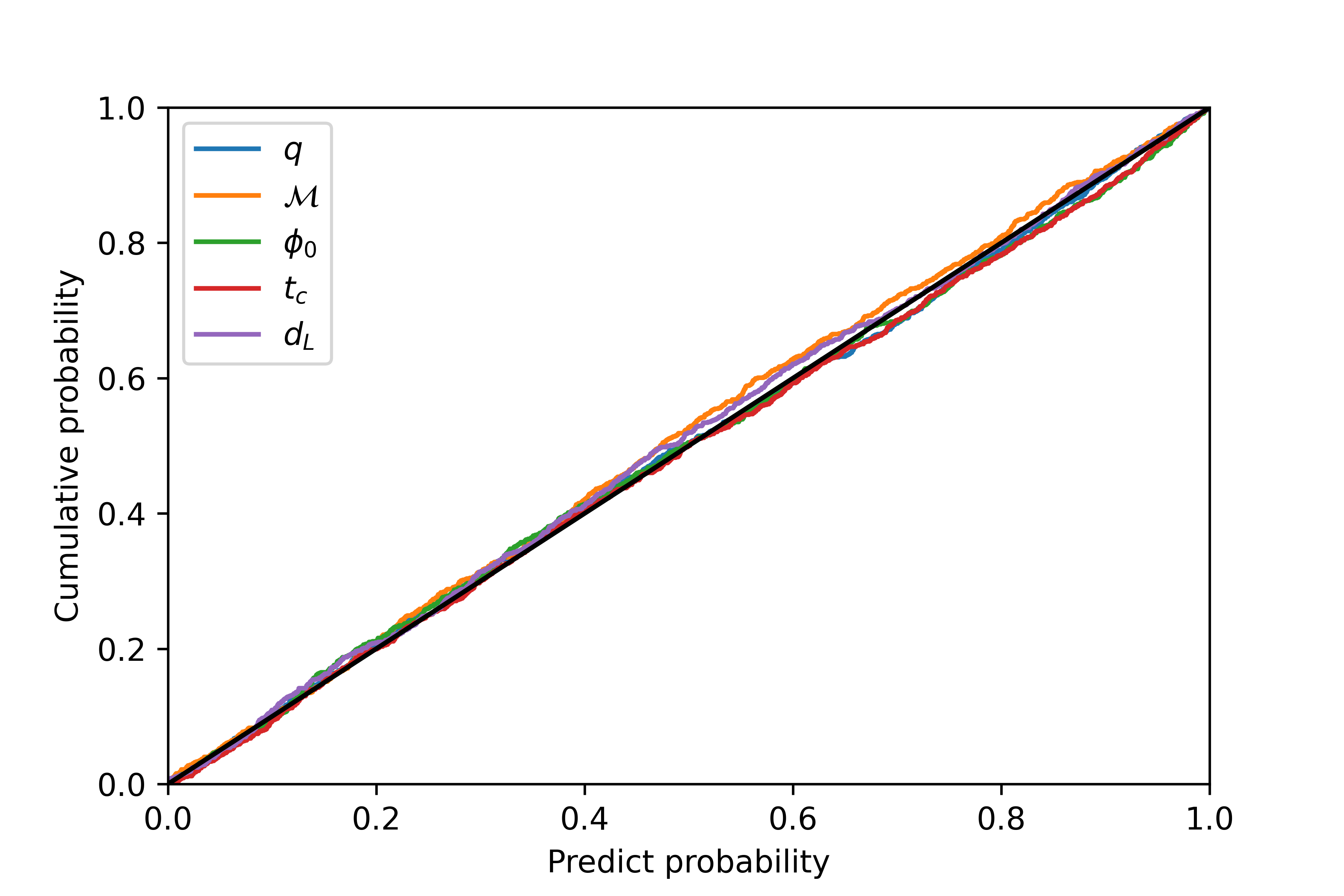}}
		\caption{P-P plot for our CVAE model. The CDF is calculated by $10^3$ mock data. For each mock data, we use $2\times 10^4$ samples to estimate the $p$-value of each parameter.}	\label{fig:pp}
\end{figure}

After training the CVAE models, the first thing is to check the self-consistency of the resultant Bayesian inference machine, i.e., calculating the P-P plot, which is the cumulative distribution function of the $p$-value of the posteriors, i.e., $p\textrm{-value}=p[p(x|y)>x|\textrm{null hypothesis}]$. By construction, the distributions of the input parameters should equal the posteriors, so that $p$-value should be the uniform of unity. Thus, the P-P plot should be diagonal for self-consistency. The result is shown in Fig. \ref{fig:pp} and indicates that our Bayesian inference machine $\overline{\textrm{CVAE}}_{\textrm{ASD}}$ is self-consistent. Compared to the P-P plot shown in Fig. 4 of \cite{gabbard2019bayesian} obtained for $\textrm{CVAE}_{\textrm{nc-ASD}}$,  the one shown here is far more convergent. This is due to the implementation of KL annealing and learning rate decay, and also to the conditional ASDs.

\begin{figure}[]
 \centering\resizebox{8cm}{!}{\includegraphics{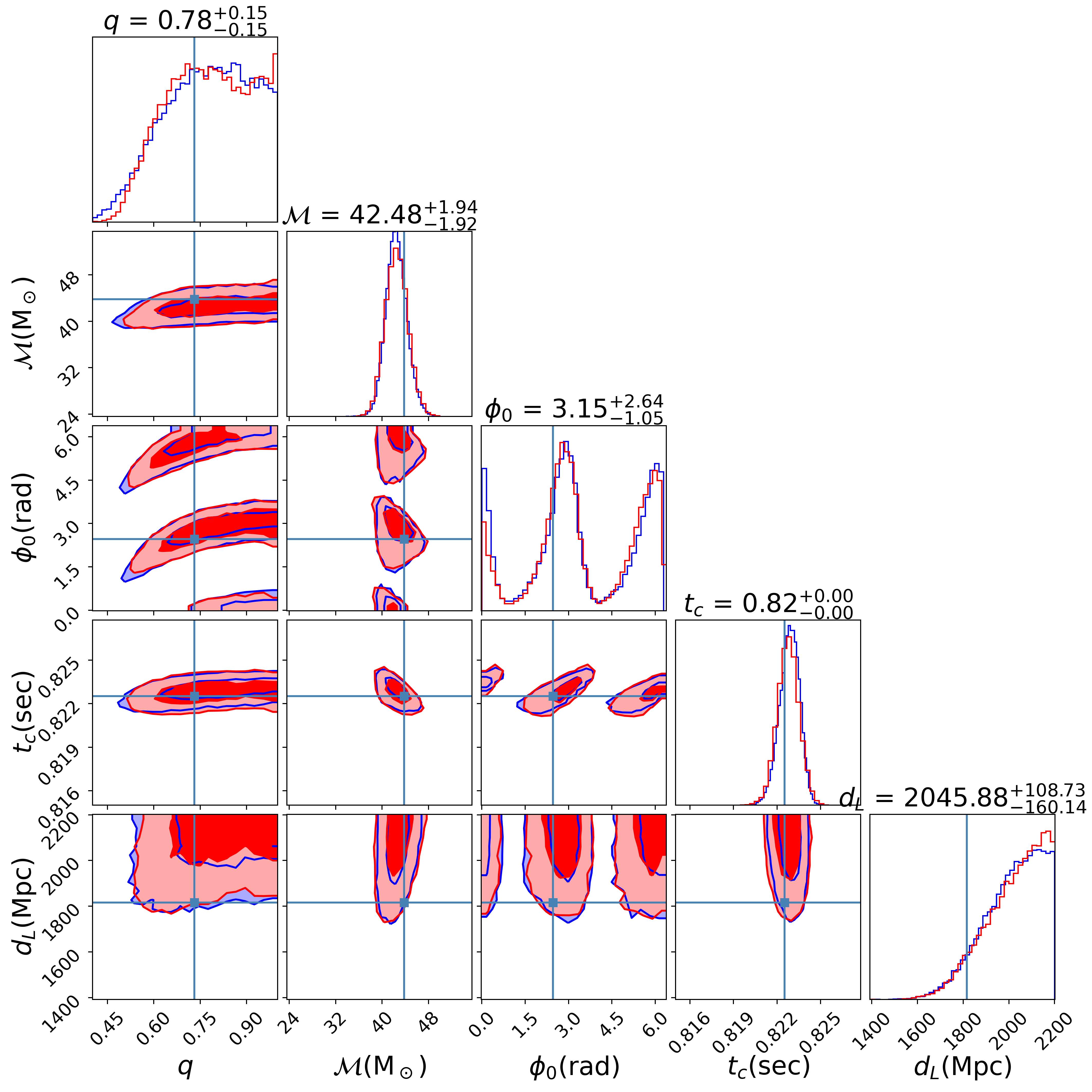}}
\caption{The marginal posteriors of a typical mock GW event evaluated from $\overline{\textrm{CVAE}}_{\textrm{ASD}}$ (red)  and the traditional PE method, i.e., dynesty (blue). 
The contour represents $50\%$ and $90\%$ credible level and the true parameter are shown by the blue lines. The KL divergences between posteriors of these two method are $(0.0005, 0.0061, 0.0184, 0.0262, 0.0010)$ in the following order of the parameters: $(q,\mathcal{M},\phi_0, t_c, d_L)$.} \label{fig:posterior}
\end{figure}

Next, we compute the posterior of a typical mock GW event by $\overline{\textrm{CVAE}}_{\textrm{ASD}}$. To produce this posterior, we need to sample about $8\times 10^4$ latent vectors from $z\sim E_{w_2}(z|y)$, and then use \eq{posterior} to average over $z$ by Monte-Carlo method to obtain the posterior $p(\theta|d, \textrm{ASD})$ for the source parameters $\theta$ \footnote{The amount of the samples, i.e., $8\times 10^4$, used here to obtain the posterior is chosen to be the same as in \cite{gabbard2019bayesian} so that we can compare the performance of our CVAE model with theirs on an  equal footing. Later on, we will also use the same amount of samples to obtain the posterior by the Nested Sampling method $\mbox{dynesty}$ for comparisons.} The results are shown in Fig. \ref{fig:posterior}, in which we also compare with the results obtained from the traditional PE algorithm dynesty. We can see that the marginal posteriors from both methods are compatible. 

One essential question about the performance of our CVAE Bayesian machine $\overline{\textrm{CVAE}}_{\textrm{ASD}}$ is how good it is when compared to $\overline{\textrm{CVAE}}_{\textrm{nc-ASD}}$. One way to characterize such a performance is to compare their KL divergences with the posterior obtained from dynesty, i.e., to compare ${\bf D}_{\textrm{KL}}(p_\textrm{dynesty}||p_\textrm{ASD})$, and ${\bf D}_{\textrm{KL}}(p_\textrm{dynesty}||p_\textrm{nc-ASD})$, where $p_\textrm{dynesty}$, $p_\textrm{ASD}$ and $p_\textrm{nc-ASD}$ denote the posteriors obtained from dynesty, $\overline{\textrm{CVAE}}_{\textrm{ASD}}$ and $\overline{\textrm{CVAE}}_{\textrm{nc-ASD}}$, respectively. Note that smaller KL divergence means the posteriors from both CVAE models are close to the one from dynesty. Usually, the  threshold for an acceptable ``nice" result is for the KL divergence to be smaller than $0.1$. Moreover, compared to the KL divergences shown in Fig. 5 of \cite{gabbard2019bayesian} obtained for $\textrm{CVAE}_{\textrm{nc-ASD}}$, the results shown here are about one to two orders better. Again, this is due to the implementation of KL annealing and learning rate decay in the training procedure. 

\begin{figure}[!htb]
\centering\resizebox{8cm}{!}{\includegraphics{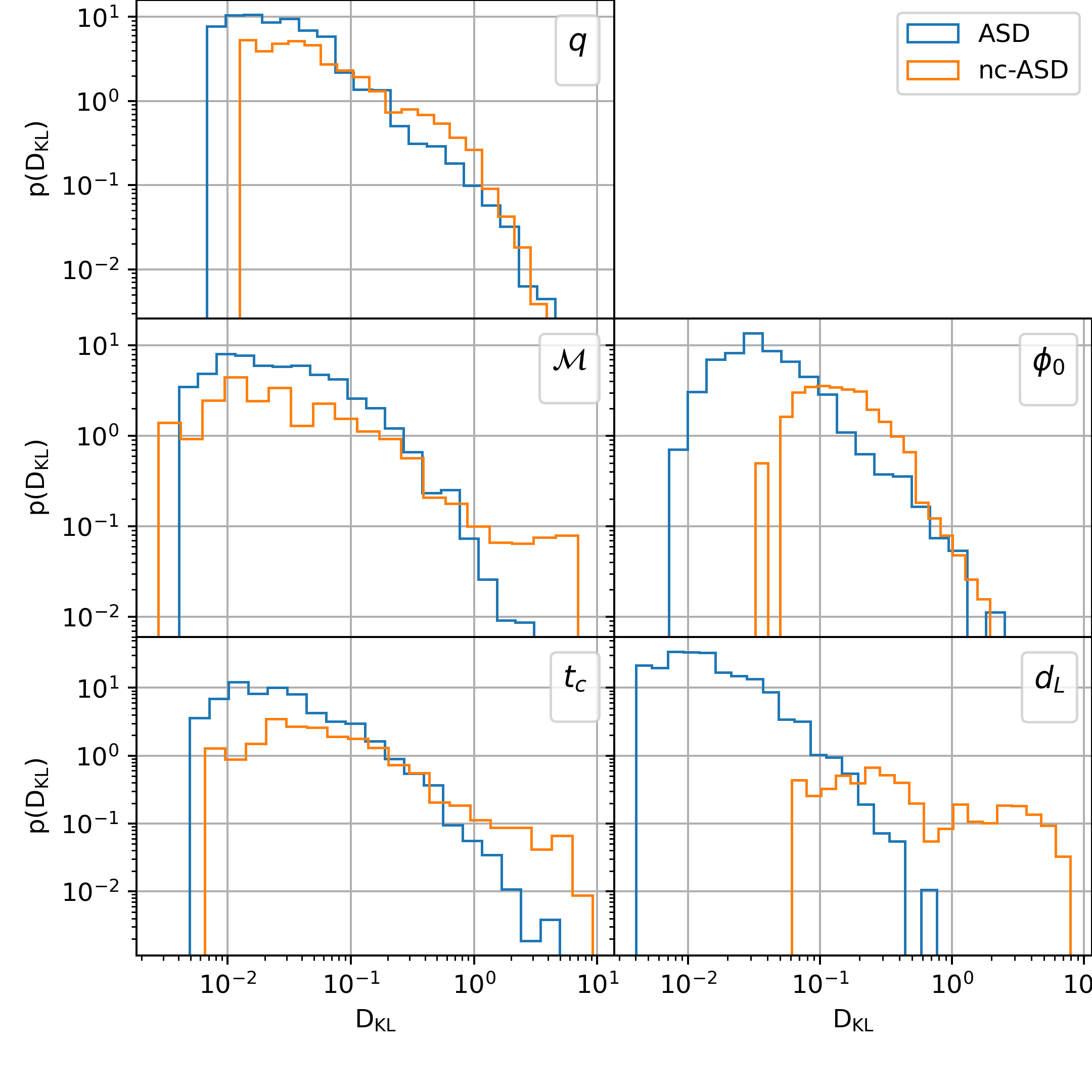}}
\caption{Histograms of KL divergences, i.e., ${\bf D}_{\textrm{KL}}(p_\textrm{dynesty}||p_\textrm{ASD})$ (blue) and ${\bf D}_{\textrm{KL}}(p_\textrm{dynesty}||p_\textrm{nc-ASD})$ (orange) for all five parameters $(q,\mathcal{M},\phi_0,t_c,d_L)$ over 512 mock GW strains of BBH with ASD variations similar to the one in Fig. \ref{fig:asd_comparison}. The preparation of these mock strains is described in the main text. Note that $p_\textrm{dynesty}$, $p_\textrm{ASD}$ and $p_\textrm{nc-ASD}$ are the posteriors obtained from dynesty, $\overline{\textrm{CVAE}}_{\textrm{ASD}}$, and $\overline{\textrm{CVAE}}_{\textrm{nc-ASD}}$, respectively. We see that $\overline{\textrm{CVAE}}_{\textrm{ASD}}$ performs better than $\overline{\textrm{CVAE}}_{\textrm{nc-ASD}}$, especially for $\phi_0$ and $d_L$ at ${\bf D}_{\textrm{KL}}\sim {\cal O}(1)$.} 	\label{fig:kld_Th}
\end{figure}

 We prepare $512$ mock GW strains as the inputs to the three Bayesian machines for comparison. These mock GW strains are generated according to the BBH priors in Table \ref{table:prior} and the ASDs obtained from \eqref{ASD_o3a}. We then evaluate the distributions of ${\bf D}_{\textrm{KL}}(p_\textrm{dynesty}||p_\textrm{ASD})$, and ${\bf D}_{\textrm{KL}}(p_\textrm{dynesty}||p_\textrm{nc-ASD})$ for all five parameters $(q, \mathcal{M}, \phi_0, t_c, d_L)$ over the above mock strains.  To obtain $p_\textrm{dynesty}$ we use Bilby to perform the dynesty sampling \cite{ashton2019bilby} with 5000 live points and dlogz equal to 0.1.  The results are shown in Fig. \ref{fig:kld_Th}. We see that $\overline{\textrm{CVAE}}_{\textrm{ASD}}$ performs better than $\overline{\textrm{CVAE}}_{\textrm{nc-ASD}}$, especially for $d_L$ at ${\bf D}_{\textrm{KL}}\sim {\cal O}(1)$ by about one order of improvement.


\begin{figure}[]
\centering\resizebox{8cm}{!}{\includegraphics{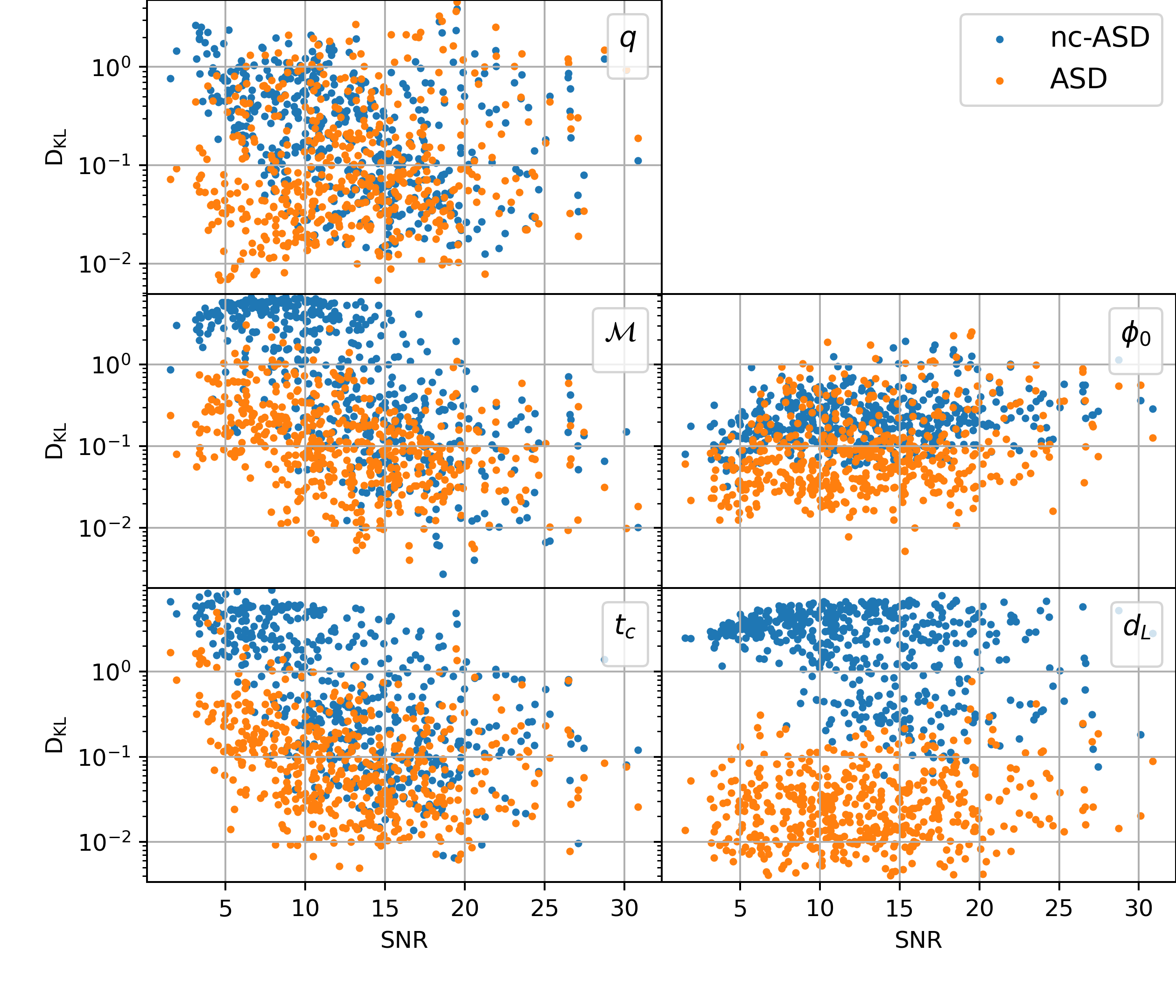}}
\caption{Comparison of the dependence on SNR for the two KL divergences for all five parameters $(q,\mathcal{M},\phi_0,t_c,d_L)$, i.e., ${\bf D}_{\textrm{KL}}[p_\textrm{dynesty}||p_\textrm{ASD}]$ (blue) and ${\bf D}_{\textrm{KL}}[p_\textrm{dynesty}||p_\textrm{nc-ASD}]$ (orange), which are already evaluated in Fig. \ref{fig:kld_Th}. We see that our CVAE model has better performance, especially at low SNR.}\label{fig:kl_injection}
\end{figure}

Besides the histograms of KL divergences shown in Fig. \ref{fig:kld_Th}, we also list these KL divergences according to the SNR of each mock event, and the result is shown in Fig. \ref{fig:kl_injection}. In general, the CVAE models perform better at higher SNR when benchmarking by the dynesty results. This could be expected for the CVAE models with primitive structures. We can also see that $\overline{\textrm{CVAE}}_{\textrm{ASD}}$ has  better overall performance than $\overline{\textrm{CVAE}}_{\textrm{nc-ASD}}$, especially for $\mathcal{M}$, $t_c$ and $d_L$, and at low SNR. The improvement of PE performance by conditioning on ASDs seems quite uniform for all SNR. This implies that our scheme can help to sort out more GW events of low SNR.


\begin{figure}[]
\centering\resizebox{8cm}{!}{\includegraphics{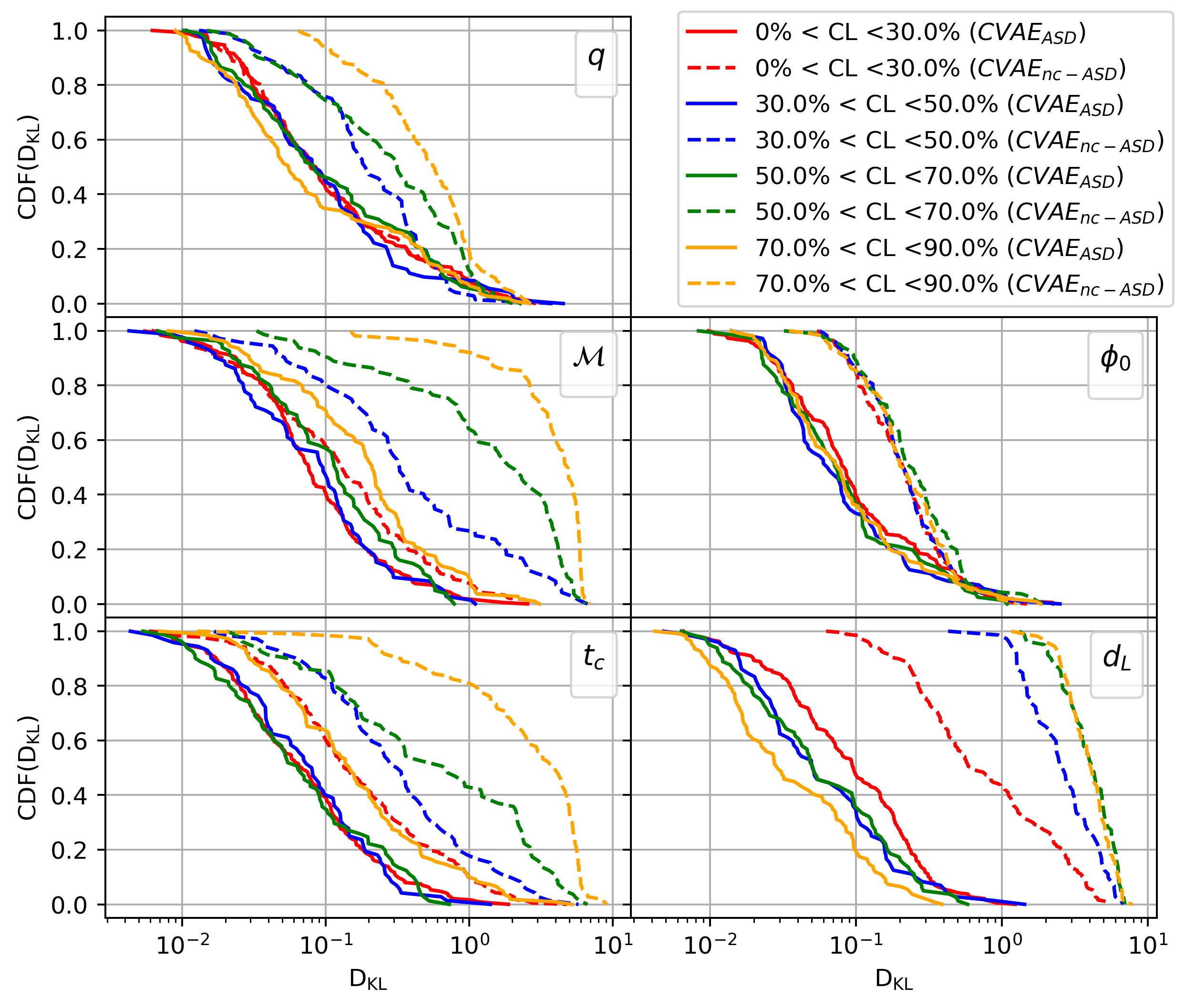}}
\caption{Comparison of the capability of $\overline{\textrm{CVAE}}_{\textrm{ASD}}$ and $\overline{\textrm{CVAE}}_{\textrm{nc-ASD}}$ in fighting against the ASD variations at various confidence levels (CLs): $0\% < \mbox{CL} < 30\%$ (red), $30\% < \mbox{CL} < 50\%$ (blue), $50\% < \mbox{CL} < 70\%$ (green) and $70\% < \mbox{CL} < 90\%$ (orange). This is characterized by plotting the cumulative distribution functions (CDFs) of KL divergences ${\bf D}_{\textrm{KL}}[p_\textrm{dynesty}||p_\textrm{ASD}]$ (solid line) and ${\bf D}_{\textrm{KL}}[p_\textrm{dynesty}||p_\textrm{nc-ASD}]$ (dashed line) of parameters $(q,\mathcal{M},\phi_0,t_c,d_L)$. The plots show that $\overline{\textrm{CVAE}}_{\textrm{ASD}}$ is better than $\overline{\textrm{CVAE}}_{\textrm{nc-ASD}}$ in fighting against ASD variations. Besides, $\overline{\textrm{CVAE}}_{\textrm{ASD}}$ is also less sensitive to the ASD variations than $\overline{\textrm{CVAE}}_{\textrm{nc-ASD}}$.} \label{fig:kld_outlier}
\end{figure}

Finally, we would like to compare the capability of  $\textrm{CVAE}_{\textrm{ASD}}$ and $\textrm{CVAE}_{\textrm{nc-ASD}}$ in fighting against ASD variations by tuning the confidence level of the overall variation factor shown in Fig. \ref{fig:asd_comparison}. To proceed, we first  classify the mock strains used in Fig. \ref{fig:kld_Th} and Fig. \ref{fig:kl_injection} by
the confidence level of $A[f]/A_m[f]$, and then plot the KL divergences ${\bf D}_{\textrm{KL}}[p_\textrm{dynesty}||p_\textrm{ASD}]$ (solid line) and ${\bf D}_{\textrm{KL}}[p_\textrm{dynesty}||p_\textrm{nc-ASD}]$ (dashed line) for each class. We show the results in Fig. \ref{fig:kld_outlier} for four different ranges of confidence level. We see that $\overline{\textrm{CVAE}}_{\textrm{ASD}}$ is always better than $\overline{\textrm{CVAE}}_{\textrm{nc-ASD}}$ in fighting against ASD variations, as expected. Especially, the PE results of $\overline{\textrm{CVAE}}_{\textrm{nc-ASD}}$ are quite sensitive to the confidence level of ASD variations, however, the ones of $\overline{\textrm{CVAE}}_{\textrm{nc-ASD}}$ are not. This indicates the superiority of our CVAE model due to conditioning on an extensive set of ASDs generated by our ASD model \eq{ASD_o3a}.

\section{Application to LIGO/Virgo's O3a events}\label{section6}

In the previous section, we have shown that our CVAE model can in general perform better in obtaining PE results for the mock events than the one without conditioning on ASD variations. We now would like to continue the similar comparison for the real LIGO/Virgo's 39 BBH O3a events, from each of which we take one-second event strain data, along with the 64-second surrounding data to construct the corresponding ASD for the standard PE practice. For simplicity, we only use the strain data from LIGO/Livingston. We choose the same priors as listed in \ref{table:prior} to perform PE of the parameters $(q,\mathcal{M},\phi_0,t_c,d_L)$ for these 39 BBH events \footnote{The prior ranges listed in Table \ref{table:prior}, especially the one for $d_L$,  are narrower than the ones used by standard LIGO/Virgo's data analysis. However, as a proof-of-concept study, we just trade the prior range/accuracy for efficiency.}. Of course, the result will depend on how well our ASD model \eq{ASD_o3a} can catch the ASD variation of these 39 BBH events with respect to $A_{m}[f]$. We will first present the PE results and the comparison with ones obtained from $\overline{\textrm{CVAE}}_{\textrm{nc-ASD}}$, and then discuss the effect of the quality of our ASD model on the results.

\begin{figure}[!htb]
\centering\resizebox{8cm}{!}{\includegraphics{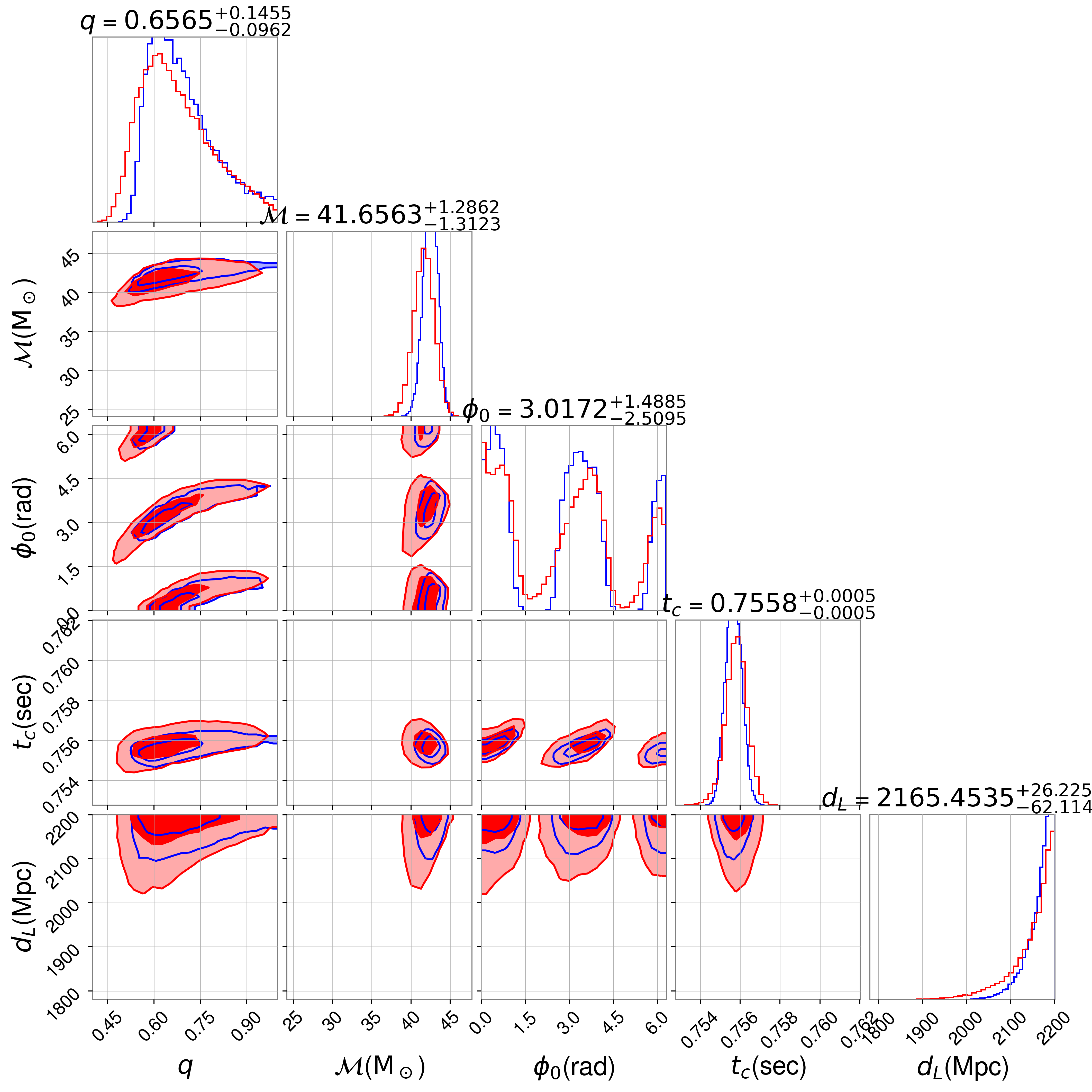}} 
\caption{Marginal posteriors of GW$190910\_112807$ event obtained by $\overline{\textrm{CVAE}}_{\textrm{ASD}}$ (red) and the dynesty (blue). The SNR of this event is $12.6$. The KL divergences, i.e.,  ${\bf D}_{\textrm{KL}}[p_\textrm{dynesty}||p_\textrm{ASD}]$, of this event for the parameters $(q, \mathcal{M}, \phi_0, t_c, d_L)$ are $(0.088, 0.29, 0.12, 0.12, 0.12 )$. It shows that the PE results obtained by $\overline{\textrm{CVAE}}_{\textrm{ASD}}$ are compatible with the ones by dynesty.  The values and the error margins of the parameters shown in this figure are the ones obtained by $\overline{\textrm{CVAE}}_{\textrm{ASD}}$, and the ones obtained by the dynesty are  $q=0.67_{-0.08}^{+0.13}$,  $\mathcal{M}=42.4_{-0.92}^{+0.90}$, $\phi_0=3.07_{-2.62}^{+1.22}$, $t_c=0.75_{-4e-4}^{+4e-4}$ and $d_L=2175.77_{-36.88}^{+17.8}$. }\label{fig:gw190910}
\end{figure}

\begin{figure}[t]
\centering\resizebox{8cm}{!}{\includegraphics{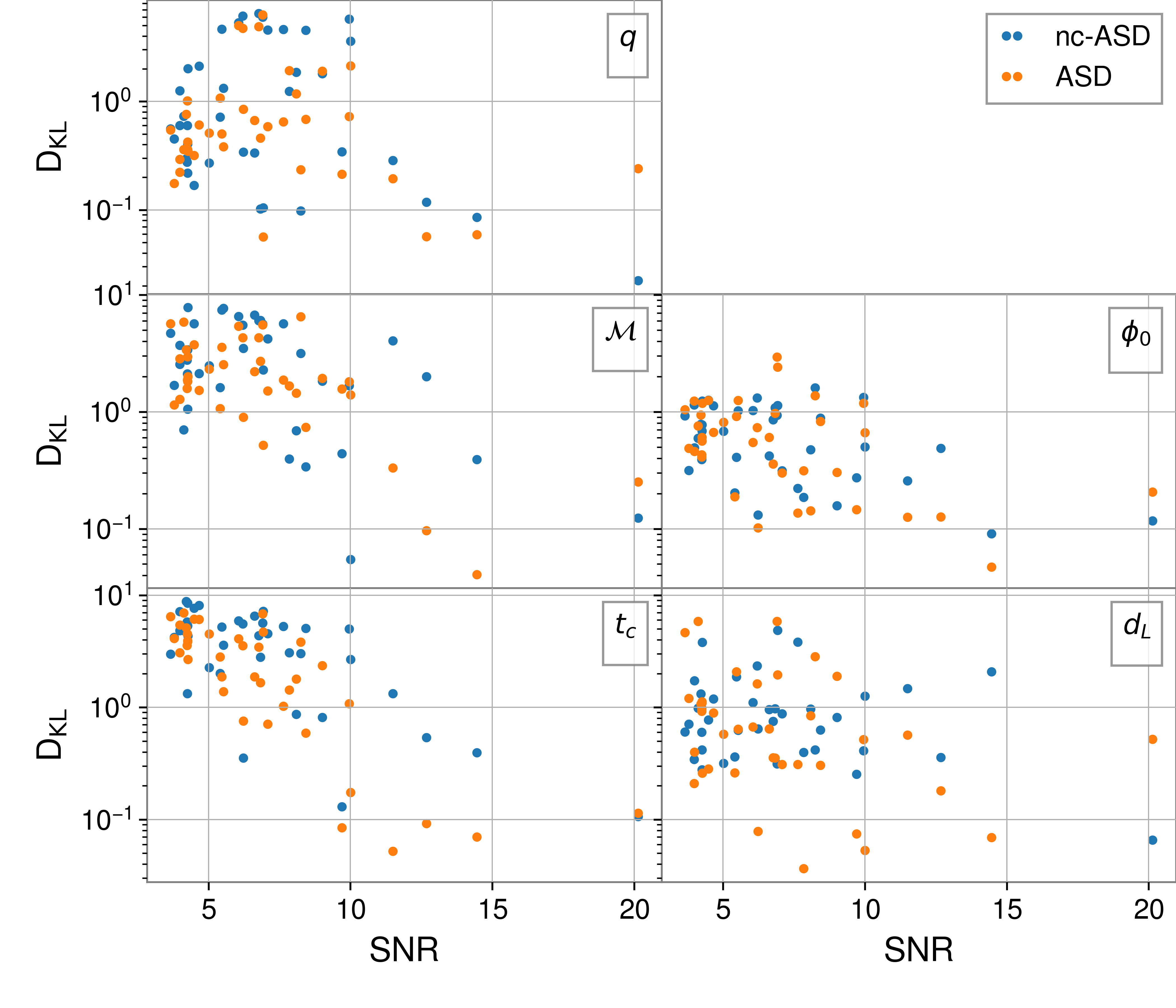}}
\caption{KL divergences, i.e., ${\bf D}_{\textrm{KL}}[p_\textrm{dynesty}||p_\textrm{ASD}]$ (orange) and ${\bf D}_{\textrm{KL}}[p_\textrm{dynesty}||p_\textrm{nc-ASD}]$ (blue) listed according to the SNRs of the 39 BBH LIGO/Virgo O3 events. The result shows that $\overline{\textrm{CVAE}}_{\textrm{ASD}}$ is better than $\overline{\textrm{CVAE}}_{\textrm{nc-ASD}}$ in the overall PE performance. However, the PE superiority of $\overline{\textrm{CVAE}}_{\textrm{ASD}}$ is not as impressive as in the cases of mock events shown in Fig. \ref{fig:kl_injection}.}\label{fig:kl_bf}
\end{figure}

In Fig. \ref{fig:gw190910} we show the marginal posteriors of the O3a event GW$190910\_112807$, of which the SNR is 12.6, one of few events with SNR larger than 10 among 39 BBH events. We see that the results obtained by $\overline{\textrm{CVAE}}_{\textrm{ASD}}$ are compatible with the ones obtained from the ones from the dynesty. This can be more precisely characterized by the values of the KL divergence for the parameters $(q,\mathcal{M},\phi_0,t_c,d_L)$, which are $(0.088, 0.29, 0.12, 0.12, 0.12 )$. The PE performance of   $\overline{\textrm{CVAE}}_{\textrm{ASD}}$ varies event by event, and the resultant KL divergences ${\bf D}_{\textrm{KL}}[p_\textrm{dynesty}||p_\textrm{ASD}]$   and ${\bf D}_{\textrm{KL}}[p_\textrm{dynesty}||p_\textrm{nc-ASD}]$ plotted according to the SNR for all 39 BBH O3a events are summarized in Fig. \ref{fig:kl_bf}. The results show that overall $\overline{\textrm{CVAE}}_{\textrm{ASD}}$ performs better than $\overline{\textrm{CVAE}}_{\textrm{nc-ASD}}$. However, the superiority of $\overline{\textrm{CVAE}}_{\textrm{ASD}}$
is not as impressive as in the case of mock events, for which the ASDs are generated simply by our ASD model \eq{ASD_o3a}. On the other hand, the key feature of the ASDs for these 39 O3a events could deviate from the ones given by \eq{ASD_o3a}.

\begin{figure}[t]
\centering\resizebox{8cm}{!}{\includegraphics{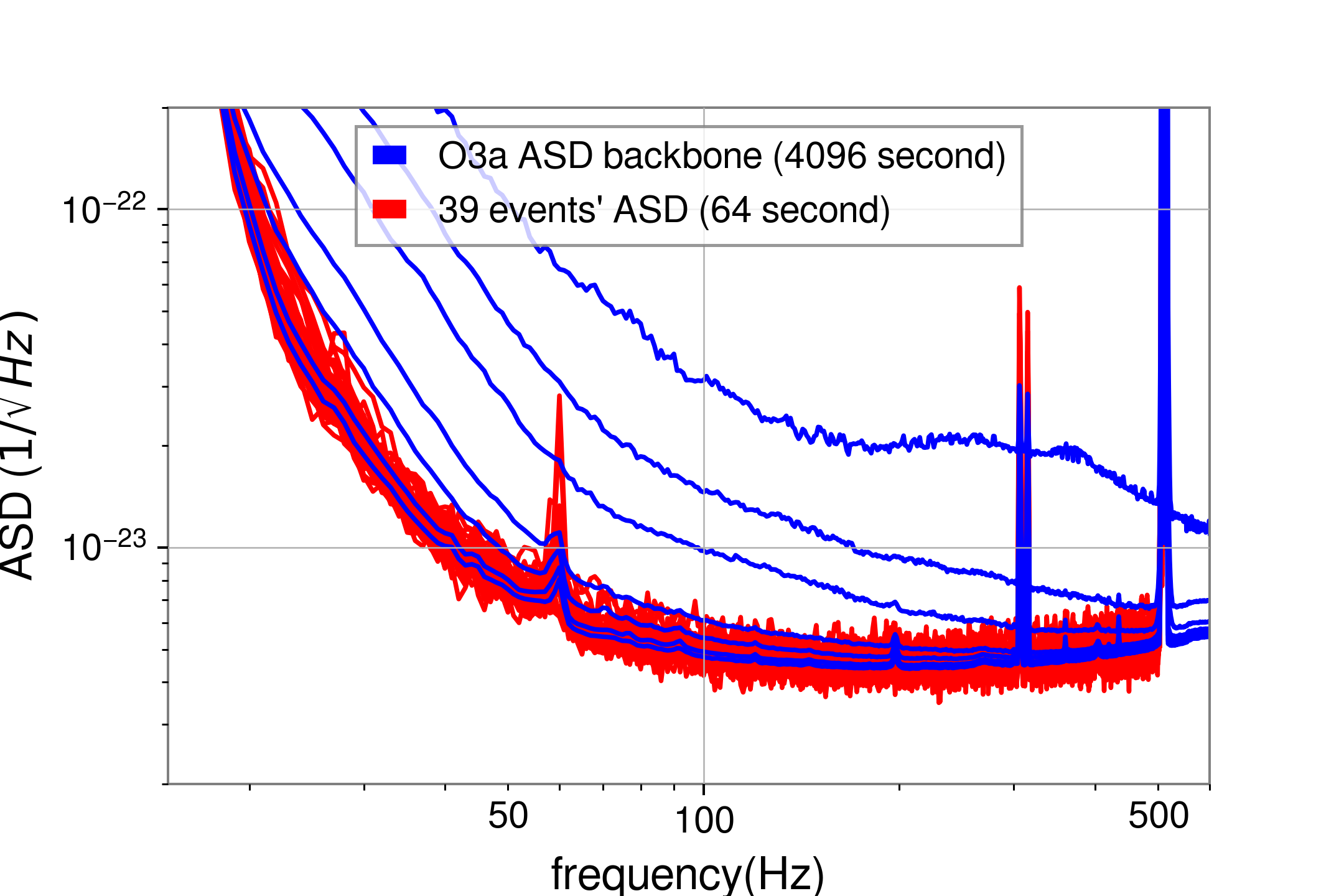}}
\caption{Comparison of the ASDs  surrounding the 39 O3a BBH events (red) and from the ASD model \eq{ASD_o3a} simulated from 2000 4096-second segments (blue). The ASD for each GW event is constructed from a surrounding 64-second segment. From the bottom to the top, the blue curves show the range of ASD variations at the confidence levels of $[1\%, 10\%, 34\%, 50\%, 76\%, 90\%, 99\%]$. }\label{fig:ASD_comp_4096}
\end{figure}

\begin{figure}[t]
\centering\resizebox{8cm}{!}{\includegraphics{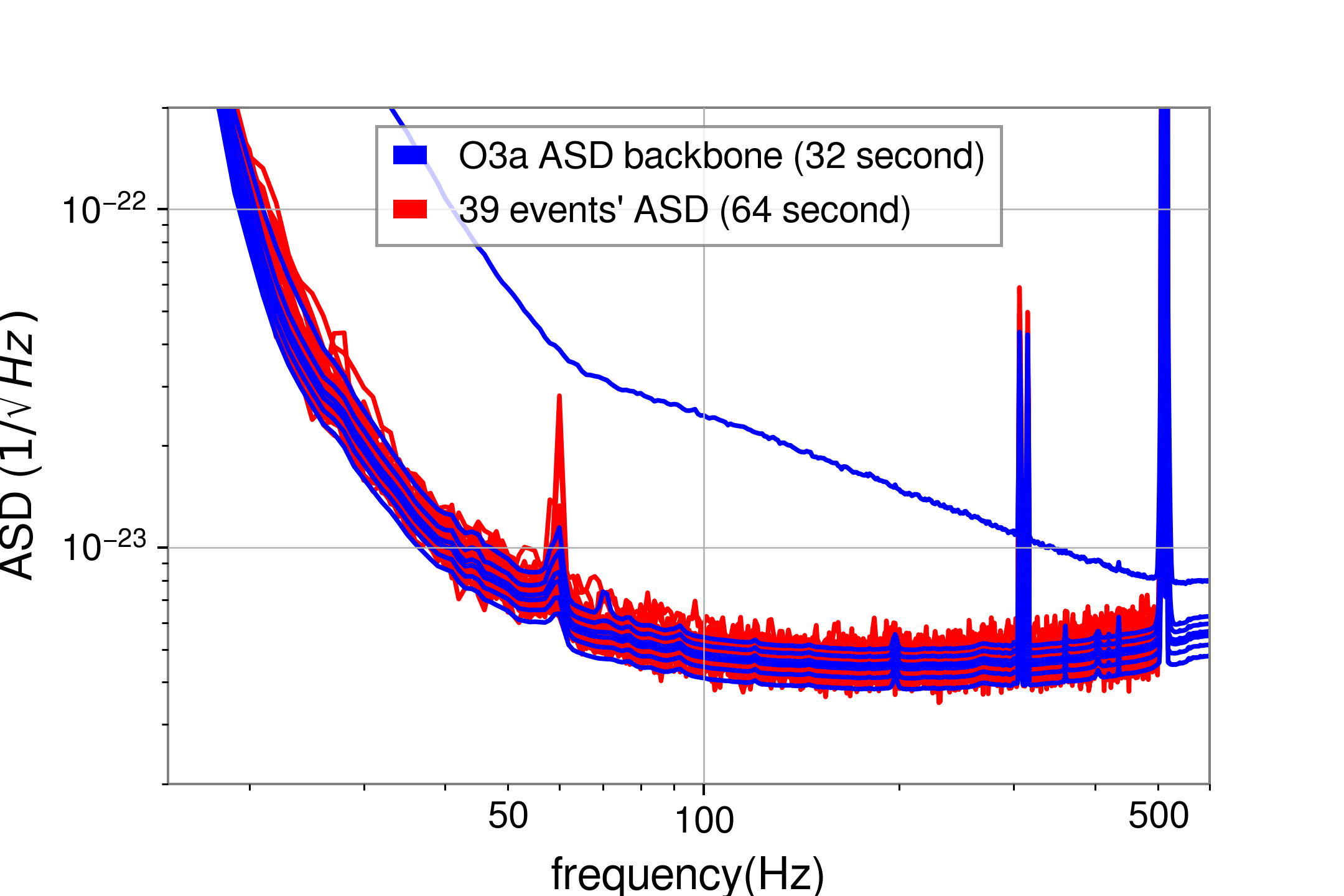}}
\caption{Similar to Fig. \ref{fig:ASD_comp_4096} but now the ASD variations are given by 16000 32-second ASDs. The legends are the same as in Fig. \ref{fig:ASD_comp_4096}. }\label{fig:ASD_comp_32}
\end{figure}

In Fig. \ref{fig:ASD_comp_4096} we compare the ASDs of the 39 O3a BBH events with the range of ASD variations from our ASD model \eq{ASD_o3a}, which is fitted by 2000 4096-second segments of O3a strain data. We can see that the ASDs of the GW events fall within the range of model ASD variations at the confidence level below $34\%$. This means that our ASD model indeed generates a far wider spectrum of ASD variations needed for the PE of these 39 GW events. It implies that we may need to enlarge our machine structure for training to suppress the unnecessary ASD uncertainty for achieving higher PE accuracy. The wider range of ASD variation can be attributed to the long 4096-second duration of the segments, which are taken to average out the glitches but then introduce large variations away from the 64-second ASD segment of the GW events. 

\begin{figure}[t]
\centering\resizebox{8cm}{!}{\includegraphics{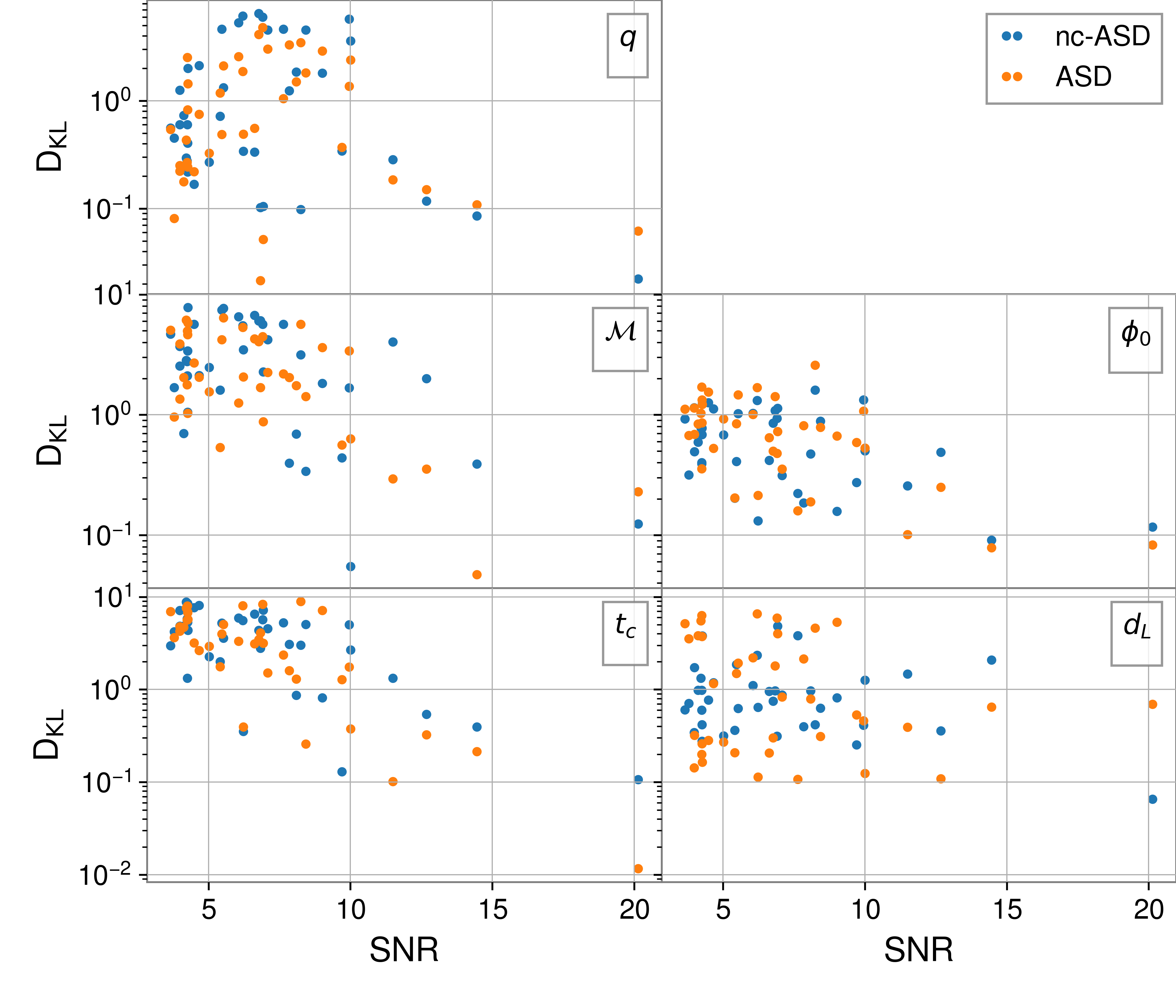}}
\caption{The PE result of our CVAE model trained by 16000 32-second ASDs. The ASD for $\overline{\textrm{CVAE}}_{\textrm{nc-ASD}}$ is just $A_m[f]$ obtained from the minimum of the collection of 32-second ASDs.  This is to compare with its counterpart trained by 4096-second ASDs as shown in Fig. \ref{fig:kl_bf}. }\label{fig:kl_snr_32}
\end{figure}

As curiosity about the effect of changing the duration of segments for the conditional ASD variations, in Fig. \ref{fig:ASD_comp_32} we show the ASD variations obtained from the collection of 16000 32-second ASDs of O3a data \footnote{In this case, we do not try to construct a stochastic ASD model like \eq{ASD_o3a} as done for the 4096-second ones. Instead, we just treat these 16000 32-second ASDs as the training set for the CVAE model. A more sophisticated ASD model based on this collection is wanted to generate more amount of variational ASDs for future improvement.}. Compared to the results in Fig. \ref{fig:ASD_comp_4096}, we can see that the new ASD distribution produces far tighter ASD variations to the 64-second ones surrounding the GW events. However, we may expect that the glitches in the ASDs may not be suppressed as much as for the 4096-second ones. Moreover, in order to compare with the 4096-second ones shown in Fig. \ref{fig:kl_bf}, in Fig. \ref{fig:kl_snr_32}  we show the PE results obtained from our CVAE models by conditioning on this collection of ASDs. On the other hand, the ASD for $\overline{\textrm{CVAE}}_{\textrm{nc-ASD}}$ is just $A_m[f]$ obtained from the minimum of this new collection of 32-second ASDs.  We find that the overall PE performances of both models are not much different. This suggests that one will face the trade-off between the long-time variations and short-time glitches when simulating an ASD model to take care of its drifting. To further improve the PE performance of our CVAE model, we may either construct a more sophisticated ASD model than the simple \eq{ASD_o3a} or enlarge the complexity of the neural network to accommodate more subtle ASD variations. Either way needs more trials and errors, such as the one done by \cite{dax2021realtime}.

\section{Conclusion} \label{conclusion}
 
In this work, we construct a deep Bayesian machine conditioning on the variations of the detector's noise to perform the parameter estimation (PE) of binary black holes' gravitational wave (GW) events based on the deep learning scheme of conditional variational autoencoder (CVAE).  This is a simple extension of the CVAE model proposed in \cite{gabbard2019bayesian} in which only strains but not the amplitude spectral density (ASD) of the detector noise are adopted as the conditional inputs to CVAE. Our motivation is to demonstrate the viability of a deep Bayesian machine that can adapt to the variations or drift of the detector's noise. This kind of machine can save time for retraining when performing PE for various GW events with slight variations of the detector noise.

As a proof of concept study, we choose a very simple layer structure, i.e., three dense layers, for two encoders and one decoder of CVAE. Despite such a humble deep machine, we show that the PE results for the mock strains with variations from a theoretical ASD are compatible with the ones obtained from the traditional PE method such as the dynesty once the tricks of KL annealing and learning-rate decay are implemented in the training procedure. Besides, we also show that our CVAE machine is better than the one of \cite{gabbard2019bayesian} in fighting against the ASD variations.

To test our CVAE model for real events and demonstrate the relevance of conditioning on the detector's noise, we also apply our CVAE Bayesian machine to 39 BBH LIGO/Virgo O3 GW events. We find that our PE performance can be compatible with the traditional Nested Sampling method for SNR larger than some threshold value, and overall is better than the CVAE model without conditioning on the ASDs. We also discuss the possible ways to further improve the PE performance of our CVAE model by constructing more sophisticated ASD models or more complicated neural networks. Finally, we hope that what we have considered in this paper can help to boost the PE tasks by deep learning for future GW events.

\section*{Acknowledgement}
This work is supported by Taiwan's Ministry of Science and Technology (MoST) through Grant No.~109-2112-M-003-007-MY3. We thank Guo-Chin Liu for generosity in providing support of the computing facility.  We also thank NCTS for partial financial support. Finally, We thank TGWG members for the helpful discussions. 

\bigskip

\appendix
\section{Fitting a stochastic ASD model for O3a run}\label{fitting}
Here we present the method about how we fix the drifting factor $r[f]$ in \eq{ASD_o3a}. We would like to simulate the ASD of O3a and its variations so that we can rewrite \eq{ASD_o3a} into:
\begin{equation}\label{ASD_log}
    \log{A_{O3a}[f]} =\alpha  \log{r[f]} + \log{A_m[f]}\;,
\end{equation}
where $A_{O3a}[f]$ is picked up from the 2000 samples of O3a ASD curves, and $r[f]$ is a function of frequency bins to be fixed. Here we neglect the residual variation $\beta$ since it introduces a small variation relative to the distribution $\alpha \log{r[f]}$. On the left side of \eq{ASD_log}, we can form a probability density functions (PDF) out of the 2000 samples of $A_{O3a}[f]$, and on the right side, it is a PDF associated with the chi-square distribution $\alpha$. The overlap between these two PDFs then depends on $\log{r[f]}$. By maximizing the overlap, we can determine $\log{r[f]}$ bin by bin. In this way, we obtain our stochastic ASD model \eq{ASD_o3a}.

The above construction has assumed the PDF of  $\log{A_{O3a}[f]}$ is nothing but a chi-square distribution, which is however justified by the fact that the maximal overlap is almost perfect, i.e., almost $100\%$. Besides, the bin-by-bin construction may also ignore the correlations between different bins, such that our ASD model  \eq{ASD_o3a} may not capture the full characteristics of O3a's ASD and its variations. Despite that, the fitted $r[f]$ shown in Fig. \ref{fig:drifting_factor} is indeed a smooth function of frequency. This suggests that the above construction is reasonable. 

\bibliography{cVAE.bib} 

\begin{thebibliography}{30}%
\makeatletter
\providecommand \@ifxundefined [1]{%
 \@ifx{#1\undefined}
}%
\providecommand \@ifnum [1]{%
 \ifnum #1\expandafter \@firstoftwo
 \else \expandafter \@secondoftwo
 \fi
}%
\providecommand \@ifx [1]{%
 \ifx #1\expandafter \@firstoftwo
 \else \expandafter \@secondoftwo
 \fi
}%
\providecommand \natexlab [1]{#1}%
\providecommand \enquote  [1]{``#1''}%
\providecommand \bibnamefont  [1]{#1}%
\providecommand \bibfnamefont [1]{#1}%
\providecommand \citenamefont [1]{#1}%
\providecommand \href@noop [0]{\@secondoftwo}%
\providecommand \href [0]{\begingroup \@sanitize@url \@href}%
\providecommand \@href[1]{\@@startlink{#1}\@@href}%
\providecommand \@@href[1]{\endgroup#1\@@endlink}%
\providecommand \@sanitize@url [0]{\catcode `\\12\catcode `\$12\catcode
  `\&12\catcode `\#12\catcode `\^12\catcode `\_12\catcode `\%12\relax}%
\providecommand \@@startlink[1]{}%
\providecommand \@@endlink[0]{}%
\providecommand \url  [0]{\begingroup\@sanitize@url \@url }%
\providecommand \@url [1]{\endgroup\@href {#1}{\urlprefix }}%
\providecommand \urlprefix  [0]{URL }%
\providecommand \Eprint [0]{\href }%
\providecommand \doibase [0]{http://dx.doi.org/}%
\providecommand \selectlanguage [0]{\@gobble}%
\providecommand \bibinfo  [0]{\@secondoftwo}%
\providecommand \bibfield  [0]{\@secondoftwo}%
\providecommand \translation [1]{[#1]}%
\providecommand \BibitemOpen [0]{}%
\providecommand \bibitemStop [0]{}%
\providecommand \bibitemNoStop [0]{.\EOS\space}%
\providecommand \EOS [0]{\spacefactor3000\relax}%
\providecommand \BibitemShut  [1]{\csname bibitem#1\endcsname}%
\let\auto@bib@innerbib\@empty
\bibitem [{\citenamefont {Abbott}\ \emph {et~al.}(2016)\citenamefont {Abbott}
  \emph {et~al.}}]{Abbott:2016blz}%
  \BibitemOpen
  \bibfield  {author} {\bibinfo {author} {\bibfnamefont {B.}~\bibnamefont
  {Abbott}} \emph {et~al.} (\bibinfo {collaboration} {LIGO Scientific,
  Virgo}),\ }\href {\doibase 10.1103/PhysRevLett.116.061102} {\bibfield
  {journal} {\bibinfo  {journal} {Phys. Rev. Lett.}\ }\textbf {\bibinfo
  {volume} {116}},\ \bibinfo {pages} {061102} (\bibinfo {year} {2016})},\
  \Eprint {http://arxiv.org/abs/1602.03837} {arXiv:1602.03837 [gr-qc]}
  \BibitemShut {NoStop}%
\bibitem [{\citenamefont {Abbott}\ \emph {et~al.}(2019)\citenamefont {Abbott}
  \emph {et~al.}}]{LIGOScientific:2018mvr}%
  \BibitemOpen
  \bibfield  {author} {\bibinfo {author} {\bibfnamefont {B.}~\bibnamefont
  {Abbott}} \emph {et~al.} (\bibinfo {collaboration} {LIGO Scientific,
  Virgo}),\ }\href {\doibase 10.1103/PhysRevX.9.031040} {\bibfield  {journal}
  {\bibinfo  {journal} {Phys. Rev. X}\ }\textbf {\bibinfo {volume} {9}},\
  \bibinfo {pages} {031040} (\bibinfo {year} {2019})},\ \Eprint
  {http://arxiv.org/abs/1811.12907} {arXiv:1811.12907 [astro-ph.HE]}
  \BibitemShut {NoStop}%
\bibitem [{\citenamefont {Abbott}\ \emph
  {et~al.}(2020{\natexlab{a}})\citenamefont {Abbott} \emph
  {et~al.}}]{Abbott:2020niy}%
  \BibitemOpen
  \bibfield  {author} {\bibinfo {author} {\bibfnamefont {R.}~\bibnamefont
  {Abbott}} \emph {et~al.} (\bibinfo {collaboration} {LIGO Scientific,
  Virgo}),\ }\href@noop {} {\  (\bibinfo {year} {2020}{\natexlab{a}})},\
  \Eprint {http://arxiv.org/abs/2010.14527} {arXiv:2010.14527 [gr-qc]}
  \BibitemShut {NoStop}%
\bibitem [{\citenamefont {Skilling}(2004)}]{skilling2004nested}%
  \BibitemOpen
  \bibfield  {author} {\bibinfo {author} {\bibfnamefont {J.}~\bibnamefont
  {Skilling}},\ }in\ \href@noop {} {\emph {\bibinfo {booktitle} {AIP Conference
  Proceedings}}},\ Vol.\ \bibinfo {volume} {735}\ (\bibinfo {organization}
  {American Institute of Physics},\ \bibinfo {year} {2004})\ pp.\ \bibinfo
  {pages} {395--405}\BibitemShut {NoStop}%
\bibitem [{\citenamefont {Del~Pozzo}\ and\ \citenamefont
  {Veitch}(2015)}]{del2015cpnest}%
  \BibitemOpen
  \bibfield  {author} {\bibinfo {author} {\bibfnamefont {W.}~\bibnamefont
  {Del~Pozzo}}\ and\ \bibinfo {author} {\bibfnamefont {J.}~\bibnamefont
  {Veitch}},\ }\href@noop {} {\bibfield  {journal} {\bibinfo  {journal} {GitHub
  https://github. com/johnveitch/cpnest}\ } (\bibinfo {year}
  {2015})}\BibitemShut {NoStop}%
\bibitem [{\citenamefont {Speagle}(2020)}]{speagle2020dynesty}%
  \BibitemOpen
  \bibfield  {author} {\bibinfo {author} {\bibfnamefont {J.~S.}\ \bibnamefont
  {Speagle}},\ }\href@noop {} {\bibfield  {journal} {\bibinfo  {journal}
  {Monthly Notices of the Royal Astronomical Society}\ }\textbf {\bibinfo
  {volume} {493}},\ \bibinfo {pages} {3132} (\bibinfo {year}
  {2020})}\BibitemShut {NoStop}%
\bibitem [{\citenamefont {Foreman-Mackey}\ \emph {et~al.}(2013)\citenamefont
  {Foreman-Mackey}, \citenamefont {Hogg}, \citenamefont {Lang},\ and\
  \citenamefont {Goodman}}]{foreman2013emcee}%
  \BibitemOpen
  \bibfield  {author} {\bibinfo {author} {\bibfnamefont {D.}~\bibnamefont
  {Foreman-Mackey}}, \bibinfo {author} {\bibfnamefont {D.~W.}\ \bibnamefont
  {Hogg}}, \bibinfo {author} {\bibfnamefont {D.}~\bibnamefont {Lang}}, \ and\
  \bibinfo {author} {\bibfnamefont {J.}~\bibnamefont {Goodman}},\ }\href@noop
  {} {\bibfield  {journal} {\bibinfo  {journal} {Publications of the
  Astronomical Society of the Pacific}\ }\textbf {\bibinfo {volume} {125}},\
  \bibinfo {pages} {306} (\bibinfo {year} {2013})}\BibitemShut {NoStop}%
\bibitem [{\citenamefont {Li}(2013)}]{Li:2013lza}%
  \BibitemOpen
  \bibfield  {author} {\bibinfo {author} {\bibfnamefont {T.~G.~F.}\
  \bibnamefont {Li}},\ }\emph {\bibinfo {title} {{Extracting Physics from
  Gravitational Waves: Testing the Strong-field Dynamics of General Relativity
  and Inferring the Large-scale Structure of the Universe}}},\ \href@noop {}
  {Ph.D. thesis},\ \bibinfo  {school} {Vrije U., Amsterdam} (\bibinfo {year}
  {2013})\BibitemShut {NoStop}%
\bibitem [{\citenamefont {Abbott}\ \emph
  {et~al.}(2020{\natexlab{b}})\citenamefont {Abbott} \emph
  {et~al.}}]{KAGRA:2020npa}%
  \BibitemOpen
  \bibfield  {author} {\bibinfo {author} {\bibfnamefont {B.~P.}\ \bibnamefont
  {Abbott}} \emph {et~al.} (\bibinfo {collaboration} {KAGRA, LIGO Scientific,
  Virgo}),\ }\href {\doibase 10.1007/s41114-020-00026-9} {\bibfield  {journal}
  {\bibinfo  {journal} {Living Rev. Rel.}\ }\textbf {\bibinfo {volume} {23}},\
  \bibinfo {pages} {3} (\bibinfo {year} {2020}{\natexlab{b}})}\BibitemShut
  {NoStop}%
\bibitem [{\citenamefont {Allen}\ \emph {et~al.}(2012)\citenamefont {Allen},
  \citenamefont {Anderson}, \citenamefont {Brady}, \citenamefont {Brown},\ and\
  \citenamefont {Creighton}}]{PhysRevD.85.122006}%
  \BibitemOpen
  \bibfield  {author} {\bibinfo {author} {\bibfnamefont {B.}~\bibnamefont
  {Allen}}, \bibinfo {author} {\bibfnamefont {W.~G.}\ \bibnamefont {Anderson}},
  \bibinfo {author} {\bibfnamefont {P.~R.}\ \bibnamefont {Brady}}, \bibinfo
  {author} {\bibfnamefont {D.~A.}\ \bibnamefont {Brown}}, \ and\ \bibinfo
  {author} {\bibfnamefont {J.~D.~E.}\ \bibnamefont {Creighton}},\ }\href
  {\doibase 10.1103/PhysRevD.85.122006} {\bibfield  {journal} {\bibinfo
  {journal} {Phys. Rev. D}\ }\textbf {\bibinfo {volume} {85}},\ \bibinfo
  {pages} {122006} (\bibinfo {year} {2012})}\BibitemShut {NoStop}%
\bibitem [{\citenamefont {Messick}\ \emph {et~al.}(2017)\citenamefont
  {Messick}, \citenamefont {Blackburn}, \citenamefont {Brady}, \citenamefont
  {Brockill}, \citenamefont {Cannon}, \citenamefont {Cariou}, \citenamefont
  {Caudill}, \citenamefont {Chamberlin}, \citenamefont {Creighton},
  \citenamefont {Everett},\ and\ \citenamefont {et~al.}}]{Messick_2017}%
  \BibitemOpen
  \bibfield  {author} {\bibinfo {author} {\bibfnamefont {C.}~\bibnamefont
  {Messick}}, \bibinfo {author} {\bibfnamefont {K.}~\bibnamefont {Blackburn}},
  \bibinfo {author} {\bibfnamefont {P.}~\bibnamefont {Brady}}, \bibinfo
  {author} {\bibfnamefont {P.}~\bibnamefont {Brockill}}, \bibinfo {author}
  {\bibfnamefont {K.}~\bibnamefont {Cannon}}, \bibinfo {author} {\bibfnamefont
  {R.}~\bibnamefont {Cariou}}, \bibinfo {author} {\bibfnamefont
  {S.}~\bibnamefont {Caudill}}, \bibinfo {author} {\bibfnamefont {S.~J.}\
  \bibnamefont {Chamberlin}}, \bibinfo {author} {\bibfnamefont {J.~D.}\
  \bibnamefont {Creighton}}, \bibinfo {author} {\bibfnamefont {R.}~\bibnamefont
  {Everett}}, \ and\ \bibinfo {author} {\bibnamefont {et~al.}},\ }\href
  {\doibase 10.1103/physrevd.95.042001} {\bibfield  {journal} {\bibinfo
  {journal} {Physical Review D}\ }\textbf {\bibinfo {volume} {95}} (\bibinfo
  {year} {2017}),\ 10.1103/physrevd.95.042001}\BibitemShut {NoStop}%
\bibitem [{\citenamefont {Veitch}\ \emph {et~al.}(2015)\citenamefont {Veitch},
  \citenamefont {Raymond}, \citenamefont {Farr}, \citenamefont {Farr},
  \citenamefont {Graff}, \citenamefont {Vitale}, \citenamefont {Aylott},
  \citenamefont {Blackburn}, \citenamefont {Christensen}, \citenamefont
  {Coughlin} \emph {et~al.}}]{veitch2015parameter}%
  \BibitemOpen
  \bibfield  {author} {\bibinfo {author} {\bibfnamefont {J.}~\bibnamefont
  {Veitch}}, \bibinfo {author} {\bibfnamefont {V.}~\bibnamefont {Raymond}},
  \bibinfo {author} {\bibfnamefont {B.}~\bibnamefont {Farr}}, \bibinfo {author}
  {\bibfnamefont {W.}~\bibnamefont {Farr}}, \bibinfo {author} {\bibfnamefont
  {P.}~\bibnamefont {Graff}}, \bibinfo {author} {\bibfnamefont
  {S.}~\bibnamefont {Vitale}}, \bibinfo {author} {\bibfnamefont
  {B.}~\bibnamefont {Aylott}}, \bibinfo {author} {\bibfnamefont
  {K.}~\bibnamefont {Blackburn}}, \bibinfo {author} {\bibfnamefont
  {N.}~\bibnamefont {Christensen}}, \bibinfo {author} {\bibfnamefont
  {M.}~\bibnamefont {Coughlin}},  \emph {et~al.},\ }\href@noop {} {\bibfield
  {journal} {\bibinfo  {journal} {Physical Review D}\ }\textbf {\bibinfo
  {volume} {91}},\ \bibinfo {pages} {042003} (\bibinfo {year}
  {2015})}\BibitemShut {NoStop}%
\bibitem [{\citenamefont {Biwer}\ \emph {et~al.}(2019)\citenamefont {Biwer},
  \citenamefont {Capano}, \citenamefont {De}, \citenamefont {Cabero},
  \citenamefont {Brown}, \citenamefont {Nitz},\ and\ \citenamefont
  {Raymond}}]{biwer2019pycbc}%
  \BibitemOpen
  \bibfield  {author} {\bibinfo {author} {\bibfnamefont {C.~M.}\ \bibnamefont
  {Biwer}}, \bibinfo {author} {\bibfnamefont {C.~D.}\ \bibnamefont {Capano}},
  \bibinfo {author} {\bibfnamefont {S.}~\bibnamefont {De}}, \bibinfo {author}
  {\bibfnamefont {M.}~\bibnamefont {Cabero}}, \bibinfo {author} {\bibfnamefont
  {D.~A.}\ \bibnamefont {Brown}}, \bibinfo {author} {\bibfnamefont {A.~H.}\
  \bibnamefont {Nitz}}, \ and\ \bibinfo {author} {\bibfnamefont
  {V.}~\bibnamefont {Raymond}},\ }\href@noop {} {\bibfield  {journal} {\bibinfo
   {journal} {Publications of the Astronomical Society of the Pacific}\
  }\textbf {\bibinfo {volume} {131}},\ \bibinfo {pages} {024503} (\bibinfo
  {year} {2019})}\BibitemShut {NoStop}%
\bibitem [{\citenamefont {Ashton}\ \emph {et~al.}(2019)\citenamefont {Ashton},
  \citenamefont {H{\"u}bner}, \citenamefont {Lasky}, \citenamefont {Talbot},
  \citenamefont {Ackley}, \citenamefont {Biscoveanu}, \citenamefont {Chu},
  \citenamefont {Divakarla}, \citenamefont {Easter}, \citenamefont {Goncharov}
  \emph {et~al.}}]{ashton2019bilby}%
  \BibitemOpen
  \bibfield  {author} {\bibinfo {author} {\bibfnamefont {G.}~\bibnamefont
  {Ashton}}, \bibinfo {author} {\bibfnamefont {M.}~\bibnamefont {H{\"u}bner}},
  \bibinfo {author} {\bibfnamefont {P.~D.}\ \bibnamefont {Lasky}}, \bibinfo
  {author} {\bibfnamefont {C.}~\bibnamefont {Talbot}}, \bibinfo {author}
  {\bibfnamefont {K.}~\bibnamefont {Ackley}}, \bibinfo {author} {\bibfnamefont
  {S.}~\bibnamefont {Biscoveanu}}, \bibinfo {author} {\bibfnamefont
  {Q.}~\bibnamefont {Chu}}, \bibinfo {author} {\bibfnamefont {A.}~\bibnamefont
  {Divakarla}}, \bibinfo {author} {\bibfnamefont {P.~J.}\ \bibnamefont
  {Easter}}, \bibinfo {author} {\bibfnamefont {B.}~\bibnamefont {Goncharov}},
  \emph {et~al.},\ }\href@noop {} {\bibfield  {journal} {\bibinfo  {journal}
  {The Astrophysical Journal Supplement Series}\ }\textbf {\bibinfo {volume}
  {241}},\ \bibinfo {pages} {27} (\bibinfo {year} {2019})}\BibitemShut
  {NoStop}%
\bibitem [{\citenamefont {Gabbard}\ \emph {et~al.}(2019)\citenamefont
  {Gabbard}, \citenamefont {Messenger}, \citenamefont {Heng}, \citenamefont
  {Tonolini},\ and\ \citenamefont {Murray-Smith}}]{gabbard2019bayesian}%
  \BibitemOpen
  \bibfield  {author} {\bibinfo {author} {\bibfnamefont {H.}~\bibnamefont
  {Gabbard}}, \bibinfo {author} {\bibfnamefont {C.}~\bibnamefont {Messenger}},
  \bibinfo {author} {\bibfnamefont {I.~S.}\ \bibnamefont {Heng}}, \bibinfo
  {author} {\bibfnamefont {F.}~\bibnamefont {Tonolini}}, \ and\ \bibinfo
  {author} {\bibfnamefont {R.}~\bibnamefont {Murray-Smith}},\ }\href@noop {}
  {\bibfield  {journal} {\bibinfo  {journal} {arXiv preprint arXiv:1909.06296}\
  } (\bibinfo {year} {2019})}\BibitemShut {NoStop}%
\bibitem [{\citenamefont {Green}\ \emph {et~al.}(2020)\citenamefont {Green},
  \citenamefont {Simpson},\ and\ \citenamefont {Gair}}]{Green_2020}%
  \BibitemOpen
  \bibfield  {author} {\bibinfo {author} {\bibfnamefont {S.~R.}\ \bibnamefont
  {Green}}, \bibinfo {author} {\bibfnamefont {C.}~\bibnamefont {Simpson}}, \
  and\ \bibinfo {author} {\bibfnamefont {J.}~\bibnamefont {Gair}},\ }\href
  {\doibase 10.1103/physrevd.102.104057} {\bibfield  {journal} {\bibinfo
  {journal} {Physical Review D}\ }\textbf {\bibinfo {volume} {102}} (\bibinfo
  {year} {2020}),\ 10.1103/physrevd.102.104057}\BibitemShut {NoStop}%
\bibitem [{\citenamefont {Kingma}\ and\ \citenamefont
  {Welling}(2019)}]{kingma2019introduction}%
  \BibitemOpen
  \bibfield  {author} {\bibinfo {author} {\bibfnamefont {D.~P.}\ \bibnamefont
  {Kingma}}\ and\ \bibinfo {author} {\bibfnamefont {M.}~\bibnamefont
  {Welling}},\ }\href@noop {} {\bibfield  {journal} {\bibinfo  {journal} {arXiv
  preprint arXiv:1906.02691}\ } (\bibinfo {year} {2019})}\BibitemShut {NoStop}%
\bibitem [{\citenamefont {Yu}(2020)}]{yu2020tutorial}%
  \BibitemOpen
  \bibfield  {author} {\bibinfo {author} {\bibfnamefont {R.}~\bibnamefont
  {Yu}},\ }\href@noop {} {\enquote {\bibinfo {title} {A tutorial on vaes: From
  bayes' rule to lossless compression},}\ } (\bibinfo {year} {2020}),\ \Eprint
  {http://arxiv.org/abs/2006.10273} {arXiv:2006.10273 [cs.LG]} \BibitemShut
  {NoStop}%
\bibitem [{\citenamefont {Kingma}\ \emph {et~al.}(2016)\citenamefont {Kingma},
  \citenamefont {Salimans},\ and\ \citenamefont
  {Welling}}]{DBLP:journals/corr/KingmaSW16}%
  \BibitemOpen
  \bibfield  {author} {\bibinfo {author} {\bibfnamefont {D.~P.}\ \bibnamefont
  {Kingma}}, \bibinfo {author} {\bibfnamefont {T.}~\bibnamefont {Salimans}}, \
  and\ \bibinfo {author} {\bibfnamefont {M.}~\bibnamefont {Welling}},\ }\href
  {http://arxiv.org/abs/1606.04934} {\bibfield  {journal} {\bibinfo  {journal}
  {CoRR}\ }\textbf {\bibinfo {volume} {abs/1606.04934}} (\bibinfo {year}
  {2016})},\ \Eprint {http://arxiv.org/abs/1606.04934} {arXiv:1606.04934}
  \BibitemShut {NoStop}%
\bibitem [{\citenamefont {Green}\ and\ \citenamefont
  {Gair}(2021)}]{green2021complete}%
  \BibitemOpen
  \bibfield  {author} {\bibinfo {author} {\bibfnamefont {S.~R.}\ \bibnamefont
  {Green}}\ and\ \bibinfo {author} {\bibfnamefont {J.}~\bibnamefont {Gair}},\
  }\href@noop {} {\bibfield  {journal} {\bibinfo  {journal} {Machine Learning:
  Science and Technology}\ }\textbf {\bibinfo {volume} {2}},\ \bibinfo {pages}
  {03LT01} (\bibinfo {year} {2021})}\BibitemShut {NoStop}%
\bibitem [{\citenamefont {Dax}\ \emph {et~al.}(2021)\citenamefont {Dax},
  \citenamefont {Green}, \citenamefont {Gair}, \citenamefont {Macke},
  \citenamefont {Buonanno},\ and\ \citenamefont
  {Schölkopf}}]{dax2021realtime}%
  \BibitemOpen
  \bibfield  {author} {\bibinfo {author} {\bibfnamefont {M.}~\bibnamefont
  {Dax}}, \bibinfo {author} {\bibfnamefont {S.~R.}\ \bibnamefont {Green}},
  \bibinfo {author} {\bibfnamefont {J.}~\bibnamefont {Gair}}, \bibinfo {author}
  {\bibfnamefont {J.~H.}\ \bibnamefont {Macke}}, \bibinfo {author}
  {\bibfnamefont {A.}~\bibnamefont {Buonanno}}, \ and\ \bibinfo {author}
  {\bibfnamefont {B.}~\bibnamefont {Schölkopf}},\ }\href@noop {} {\enquote
  {\bibinfo {title} {Real-time gravitational-wave science with neural posterior
  estimation},}\ } (\bibinfo {year} {2021}),\ \Eprint
  {http://arxiv.org/abs/2106.12594} {arXiv:2106.12594 [gr-qc]} \BibitemShut
  {NoStop}%
\bibitem [{\citenamefont {Abadi}\ \emph {et~al.}(2015)\citenamefont {Abadi},
  \citenamefont {Agarwal}, \citenamefont {Barham}, \citenamefont {Brevdo},
  \citenamefont {Chen}, \citenamefont {Citro}, \citenamefont {Corrado},
  \citenamefont {Davis}, \citenamefont {Dean}, \citenamefont {Devin},
  \citenamefont {Ghemawat}, \citenamefont {Goodfellow}, \citenamefont {Harp},
  \citenamefont {Irving}, \citenamefont {Isard}, \citenamefont {Jia},
  \citenamefont {Jozefowicz}, \citenamefont {Kaiser}, \citenamefont {Kudlur},
  \citenamefont {Levenberg}, \citenamefont {Man\'{e}}, \citenamefont {Monga},
  \citenamefont {Moore}, \citenamefont {Murray}, \citenamefont {Olah},
  \citenamefont {Schuster}, \citenamefont {Shlens}, \citenamefont {Steiner},
  \citenamefont {Sutskever}, \citenamefont {Talwar}, \citenamefont {Tucker},
  \citenamefont {Vanhoucke}, \citenamefont {Vasudevan}, \citenamefont
  {Vi\'{e}gas}, \citenamefont {Vinyals}, \citenamefont {Warden}, \citenamefont
  {Wattenberg}, \citenamefont {Wicke}, \citenamefont {Yu},\ and\ \citenamefont
  {Zheng}}]{tensorflow2015-whitepaper}%
  \BibitemOpen
  \bibfield  {author} {\bibinfo {author} {\bibfnamefont {M.}~\bibnamefont
  {Abadi}}, \bibinfo {author} {\bibfnamefont {A.}~\bibnamefont {Agarwal}},
  \bibinfo {author} {\bibfnamefont {P.}~\bibnamefont {Barham}}, \bibinfo
  {author} {\bibfnamefont {E.}~\bibnamefont {Brevdo}}, \bibinfo {author}
  {\bibfnamefont {Z.}~\bibnamefont {Chen}}, \bibinfo {author} {\bibfnamefont
  {C.}~\bibnamefont {Citro}}, \bibinfo {author} {\bibfnamefont {G.~S.}\
  \bibnamefont {Corrado}}, \bibinfo {author} {\bibfnamefont {A.}~\bibnamefont
  {Davis}}, \bibinfo {author} {\bibfnamefont {J.}~\bibnamefont {Dean}},
  \bibinfo {author} {\bibfnamefont {M.}~\bibnamefont {Devin}}, \bibinfo
  {author} {\bibfnamefont {S.}~\bibnamefont {Ghemawat}}, \bibinfo {author}
  {\bibfnamefont {I.}~\bibnamefont {Goodfellow}}, \bibinfo {author}
  {\bibfnamefont {A.}~\bibnamefont {Harp}}, \bibinfo {author} {\bibfnamefont
  {G.}~\bibnamefont {Irving}}, \bibinfo {author} {\bibfnamefont
  {M.}~\bibnamefont {Isard}}, \bibinfo {author} {\bibfnamefont
  {Y.}~\bibnamefont {Jia}}, \bibinfo {author} {\bibfnamefont {R.}~\bibnamefont
  {Jozefowicz}}, \bibinfo {author} {\bibfnamefont {L.}~\bibnamefont {Kaiser}},
  \bibinfo {author} {\bibfnamefont {M.}~\bibnamefont {Kudlur}}, \bibinfo
  {author} {\bibfnamefont {J.}~\bibnamefont {Levenberg}}, \bibinfo {author}
  {\bibfnamefont {D.}~\bibnamefont {Man\'{e}}}, \bibinfo {author}
  {\bibfnamefont {R.}~\bibnamefont {Monga}}, \bibinfo {author} {\bibfnamefont
  {S.}~\bibnamefont {Moore}}, \bibinfo {author} {\bibfnamefont
  {D.}~\bibnamefont {Murray}}, \bibinfo {author} {\bibfnamefont
  {C.}~\bibnamefont {Olah}}, \bibinfo {author} {\bibfnamefont {M.}~\bibnamefont
  {Schuster}}, \bibinfo {author} {\bibfnamefont {J.}~\bibnamefont {Shlens}},
  \bibinfo {author} {\bibfnamefont {B.}~\bibnamefont {Steiner}}, \bibinfo
  {author} {\bibfnamefont {I.}~\bibnamefont {Sutskever}}, \bibinfo {author}
  {\bibfnamefont {K.}~\bibnamefont {Talwar}}, \bibinfo {author} {\bibfnamefont
  {P.}~\bibnamefont {Tucker}}, \bibinfo {author} {\bibfnamefont
  {V.}~\bibnamefont {Vanhoucke}}, \bibinfo {author} {\bibfnamefont
  {V.}~\bibnamefont {Vasudevan}}, \bibinfo {author} {\bibfnamefont
  {F.}~\bibnamefont {Vi\'{e}gas}}, \bibinfo {author} {\bibfnamefont
  {O.}~\bibnamefont {Vinyals}}, \bibinfo {author} {\bibfnamefont
  {P.}~\bibnamefont {Warden}}, \bibinfo {author} {\bibfnamefont
  {M.}~\bibnamefont {Wattenberg}}, \bibinfo {author} {\bibfnamefont
  {M.}~\bibnamefont {Wicke}}, \bibinfo {author} {\bibfnamefont
  {Y.}~\bibnamefont {Yu}}, \ and\ \bibinfo {author} {\bibfnamefont
  {X.}~\bibnamefont {Zheng}},\ }\href {https://www.tensorflow.org/} {\enquote
  {\bibinfo {title} {{TensorFlow}: Large-scale machine learning on
  heterogeneous systems},}\ } (\bibinfo {year} {2015}),\ \bibinfo {note}
  {software available from tensorflow.org}\BibitemShut {NoStop}%
\bibitem [{\citenamefont {Vallisneri}\ \emph {et~al.}(2015)\citenamefont
  {Vallisneri}, \citenamefont {Kanner}, \citenamefont {Williams}, \citenamefont
  {Weinstein},\ and\ \citenamefont {Stephens}}]{vallisneri2015ligo}%
  \BibitemOpen
  \bibfield  {author} {\bibinfo {author} {\bibfnamefont {M.}~\bibnamefont
  {Vallisneri}}, \bibinfo {author} {\bibfnamefont {J.}~\bibnamefont {Kanner}},
  \bibinfo {author} {\bibfnamefont {R.}~\bibnamefont {Williams}}, \bibinfo
  {author} {\bibfnamefont {A.}~\bibnamefont {Weinstein}}, \ and\ \bibinfo
  {author} {\bibfnamefont {B.}~\bibnamefont {Stephens}},\ }in\ \href@noop {}
  {\emph {\bibinfo {booktitle} {Journal of Physics: Conference Series}}},\
  Vol.\ \bibinfo {volume} {610}\ (\bibinfo {organization} {IOP Publishing},\
  \bibinfo {year} {2015})\ p.\ \bibinfo {pages} {012021}\BibitemShut {NoStop}%
\bibitem [{gra()}]{gracedbO3}%
  \BibitemOpen
  \href {https://gracedb.ligo.org/superevents/public/O3/} {\enquote {\bibinfo
  {title} {Gracedb gravitational-wave candidate event database (ligo/virgo o3
  public alerts).\url{https://gracedb.ligo.org/superevents/public/O3/}},}\
  }\BibitemShut {NoStop}%
\bibitem [{\citenamefont {Khan}\ \emph {et~al.}(2019)\citenamefont {Khan},
  \citenamefont {Chatziioannou}, \citenamefont {Hannam},\ and\ \citenamefont
  {Ohme}}]{khan2019phenomenological}%
  \BibitemOpen
  \bibfield  {author} {\bibinfo {author} {\bibfnamefont {S.}~\bibnamefont
  {Khan}}, \bibinfo {author} {\bibfnamefont {K.}~\bibnamefont {Chatziioannou}},
  \bibinfo {author} {\bibfnamefont {M.}~\bibnamefont {Hannam}}, \ and\ \bibinfo
  {author} {\bibfnamefont {F.}~\bibnamefont {Ohme}},\ }\href@noop {} {\bibfield
   {journal} {\bibinfo  {journal} {Physical Review D}\ }\textbf {\bibinfo
  {volume} {100}},\ \bibinfo {pages} {024059} (\bibinfo {year}
  {2019})}\BibitemShut {NoStop}%
\bibitem [{\citenamefont {Kingma}\ and\ \citenamefont
  {Ba}(2014)}]{kingma2014adam}%
  \BibitemOpen
  \bibfield  {author} {\bibinfo {author} {\bibfnamefont {D.~P.}\ \bibnamefont
  {Kingma}}\ and\ \bibinfo {author} {\bibfnamefont {J.}~\bibnamefont {Ba}},\
  }\href@noop {} {\bibfield  {journal} {\bibinfo  {journal} {arXiv preprint
  arXiv:1412.6980}\ } (\bibinfo {year} {2014})}\BibitemShut {NoStop}%
\bibitem [{\citenamefont {Lucas}\ \emph {et~al.}(2019)\citenamefont {Lucas},
  \citenamefont {Tucker}, \citenamefont {Grosse},\ and\ \citenamefont
  {Norouzi}}]{lucas2019understanding}%
  \BibitemOpen
  \bibfield  {author} {\bibinfo {author} {\bibfnamefont {J.}~\bibnamefont
  {Lucas}}, \bibinfo {author} {\bibfnamefont {G.}~\bibnamefont {Tucker}},
  \bibinfo {author} {\bibfnamefont {R.}~\bibnamefont {Grosse}}, \ and\ \bibinfo
  {author} {\bibfnamefont {M.}~\bibnamefont {Norouzi}},\ }\href@noop {} {\
  (\bibinfo {year} {2019})}\BibitemShut {NoStop}%
\bibitem [{\citenamefont {Fu}\ \emph {et~al.}(2019)\citenamefont {Fu},
  \citenamefont {Li}, \citenamefont {Liu}, \citenamefont {Gao}, \citenamefont
  {Celikyilmaz},\ and\ \citenamefont {Carin}}]{fu2019cyclical}%
  \BibitemOpen
  \bibfield  {author} {\bibinfo {author} {\bibfnamefont {H.}~\bibnamefont
  {Fu}}, \bibinfo {author} {\bibfnamefont {C.}~\bibnamefont {Li}}, \bibinfo
  {author} {\bibfnamefont {X.}~\bibnamefont {Liu}}, \bibinfo {author}
  {\bibfnamefont {J.}~\bibnamefont {Gao}}, \bibinfo {author} {\bibfnamefont
  {A.}~\bibnamefont {Celikyilmaz}}, \ and\ \bibinfo {author} {\bibfnamefont
  {L.}~\bibnamefont {Carin}},\ }\href@noop {} {\bibfield  {journal} {\bibinfo
  {journal} {arXiv preprint arXiv:1903.10145}\ } (\bibinfo {year}
  {2019})}\BibitemShut {NoStop}%
\bibitem [{\citenamefont {Ghasemi}\ and\ \citenamefont
  {Zahediasl}(2012)}]{ghasemi2012normality}%
  \BibitemOpen
  \bibfield  {author} {\bibinfo {author} {\bibfnamefont {A.}~\bibnamefont
  {Ghasemi}}\ and\ \bibinfo {author} {\bibfnamefont {S.}~\bibnamefont
  {Zahediasl}},\ }\href@noop {} {\bibfield  {journal} {\bibinfo  {journal}
  {International journal of endocrinology and metabolism}\ }\textbf {\bibinfo
  {volume} {10}},\ \bibinfo {pages} {486} (\bibinfo {year} {2012})}\BibitemShut
  {NoStop}%
\bibitem [{\citenamefont {P{\'e}rez-Cruz}(2008)}]{perez2008kullback}%
  \BibitemOpen
  \bibfield  {author} {\bibinfo {author} {\bibfnamefont {F.}~\bibnamefont
  {P{\'e}rez-Cruz}},\ }in\ \href@noop {} {\emph {\bibinfo {booktitle} {2008
  IEEE international symposium on information theory}}}\ (\bibinfo
  {organization} {IEEE},\ \bibinfo {year} {2008})\ pp.\ \bibinfo {pages}
  {1666--1670}\BibitemShut {NoStop}%
\end{thebibliography}%

\end{document}